\documentclass[preprint,notoc]{JHEP3}
\usepackage{epsfig}
\usepackage{cite}

\def\Dsl{\not\!\! D}

\def\to{\rightarrow}

\def\bi{\begin{itemize}}
\def\ei{\end{itemize}}

\def\c1p{C1^\prime}
\def\ta{\tilde a}
\def\tG{\widetilde G}

\def\ta{\tilde a}

\def\tg{\tilde g}

\def\tz{\widetilde Z}

\def\alt{\lesssim}
\def\agt{\gtrsim}
\def\be{\begin{equation}}  
\def\ee{\end{equation}}  
\def\bea{\begin{eqnarray}}  
\def\eea{\end{eqnarray}}  

\newcommand\njp[3]{{\it New\ J.\ Phys.\ }{\bf #1} (#2) #3}
\newcommand\jpg[3]{{\it J.\ Phys.\ G}{\bf #1} (#2) #3}
\newcommand\ijmpe[3]{{\it Int.\ J.\ Mod. Phys.\ }{\bf E#1} (#2) #3}
\newcommand\annp[3]{{\it Annals\ Phys.\ }{\bf #1} (#2) #3}
\newcommand\sjp[3]{{\it Sov.\ J.\ Nucl.\ }{\bf #1} (#2) #3}

\title{Thermal leptogenesis and the gravitino problem in the \\
Asaka-Yanagida  axion/axino dark matter scenario}

\author{Howard Baer$^{a}$, Sabine Kraml$^b$, Andre Lessa$^{a}$ and 
Sezen Sekmen$^c$\\
$^a$Dept.\ of Physics and Astronomy, University of Oklahoma, Norman, OK 73019, USA\\
$^b$Laboratoire de Physique Subatomique et de Cosmologie, UJF Grenoble 1, 
CNRS/IN2P3, INPG, 53 Avenue des Martyrs, F-38026 Grenoble, France\\
$^c$Dept.\ of Physics, Florida State University, Tallahassee, FL 32306, USA\\
E-mail: \email{baer@nhn.ou.edu}, \email{sabine.kraml@lpsc.in2p3.fr}, 
\email{lessa.a.p@gmail.com}, \email{sezen.sekmen@cern.ch}}

\abstract{
A successful implementation of thermal leptogenesis requires the re-heat
temperature after inflation $T_R$ to exceed $\sim 2\times 10^9$~GeV. 
Such a high $T_R$ value typically leads to an overproduction of gravitinos in the early universe, 
which will cause conflicts, mainly with BBN constraints. 
Asaka and Yanagida (AY) have proposed that these two issues can be reconciled
in the context of the Peccei-Quinn augmented MSSM (PQMSSM) if one adopts a 
mass hierarchy $m({\rm sparticle})>m({\rm gravitino})>m({\rm axino})$, 
with $m({\rm axino})\sim$~keV. 
In this case, sparticle decays 
bypass the gravitino, and decay more quickly to the axino LSP, thus avoiding the BBN constraints.
In addition, thermally produced gravitinos decay inertly to axion+axino, also avoiding BBN
constraints. 
We calculate the relic abundance of mixed axion/axino dark matter
in the AY scenario, and investigate under what conditions a value of $T_R$
sufficient for thermal leptogenesis can be generated. 
A high value of PQ breaking scale $f_a$ is needed to suppress overproduction of axinos, while
a small vacuum misalignment angle $\theta_i$ is needed to suppress overproduction of axions.
The large value of $f_a$ results in late decaying neutralinos.
We show that, to avoid BBN constraints, the AY scenario requires a rather low 
thermal abundance of neutralinos, 
while higher values of neutralino mass also help. 
We combine these constraint calculations along with entropy production from
late decaying saxions, and find the saxion needs to be typically 
at least several times 
heavier than the gravitino.
A successful implementation of the AY scenario suggests that LHC should discover a spectrum of SUSY particles
consistent with weak scale supergravity; that the apparent neutralino abundance is low;  
that a possible axion detection signal (probably with $m_a$ in the sub-$\mu$eV range) exists, 
but no direct or indirect signals for WIMP dark matter should be observed.
}

\begin{document}

%========================================================================================
\section{Introduction}
\label{sec:intro}
%========================================================================================
A wide assortment of data from atmospheric, solar, reactor and accelerator
experiments can all be explained in terms of
massive neutrinos with large mixing angles which undergo flavor
oscillations upon propagation through matter or the vacuum\cite{nu_review}.
The emerging picture of the physics behind neutrino oscillation data is 
most elegantly explained by the presence of massive gauge singlet 
right-hand Majorana neutrino states $N_i$ ($i=1-3$ a generation index) 
which give rise to see-saw neutrino masses\cite{seesaw}: 
$m_{\nu_i}\simeq (f_{\nu_i}v)^2/M_{N_i}$ with $f_{\nu_i}$ the neutrino Yukawa 
coupling, $v$ the vev of the Higgs field, and $M_{N_i}\sim 10^9-10^{15}$~GeV. 

In addition to explaining neutrino oscillation data, the presence of massive $N_i$ states
offers an elegant explanation of baryogenesis in terms of leptogenesis~\cite{lepto_review}, 
wherein the states $N_i$ exist in thermal equilibrium in the early universe, 
but decay asymmetrically to leptons versus anti-leptons. The lepton-anti-lepton
asymmetry is then converted to a baryon asymmetry via $B$ and $L$
violating, but $B-L$ conserving, sphaleron effects\cite{krs}. 
To realize the thermal leptogenesis scenario, the lightest of the
heavy neutrino masses $M_1$ must be $\agt 2\times 10^9$ GeV. 
In order to produce such states thermally, 
a re-heat temperature of the universe after inflation of
$T_R\agt M_1 > 2\times 10^9$~GeV is required\cite{T_R}.

Augmenting the Standard Model with a new, extremely high energy 
scale $M_{N_i}$ naturally leads to severe quadratic divergences in the 
Higgs sector which will need to be highly fine-tuned. The solution here is
to also incorporate supersymmetry (SUSY), which reduces quadratic 
divergences to merely logarithmic, and ameliorates the fine-tuning problem\cite{wss}.
While the addition of weak scale softly broken SUSY into the 
theory is actually supported by the measured values of the gauge 
couplings from LEP experiments, it also leads to new conundrums such as the gravitino problem: 
the production of gravitinos in the early universe
can lead to {\it (i)}~overproduction of LSP dark matter ({\it e.g.} the lightest
neutralino) beyond relic density limits obtained from WMAP and other
experiments, or {\it (ii)}~disruption of the successful explanation of
Big Bang nucleosynthesis by introducing late decaying quasi-stable
particles whose decay products can break up the newly minted light 
elements. The common solution to the gravitino problem\cite{gravprob} is to 
require a sufficiently low re-heat temperature such that thermal gravitino production
is suppressed enough to avoid overproduction of dark matter or
disruption of BBN\cite{kl}. For gravitino masses in the few TeV or below range,
a value of $T_R\alt 10^5$~GeV is required. Naively, this is in obvious
conflict with the $T_R$ requirements of thermal leptogenesis.

A variety of solutions have been proposed to reconcile 
leptogenesis with the SUSY gravitino problem. One is to abandon 
the ``thermal'' aspect of leptogenesis, and invoke non-thermal
leptogenesis wherein the heavy neutrino states are produced via
some scalar field decay, for instance the inflaton\cite{ntlepto}.
Another suggestion is to invoke the gravitino as LSP, so it does not decay. 
However, the gravitino LSP scenarios fall back into the BBN problem since then the NLSP
SUSY particle suffers a late decay into gravitino plus SM states
which again injects high energy particles into the post-BBN plasma.
One solution is to speed up NLSP decay via a small component
of $R$-parity violation\cite{covi,covi2}. 

In a recent work\cite{Baer:2010kw}, we proposed an alternative scenario, 
invoking mixed axion/axino dark matter, as would occur
in the Peccei-Quinn\cite{pq,ww,ksvz,dfsz} augmented MSSM 
(the PQMSSM)\cite{pqmssm,axino}. 
In this case, we invoked models with very heavy gravitinos,  
$m_{\tG}\agt 10$~TeV, so that gravitinos decay before the onset of BBN. 
Then, overproduction of dark matter can be avoided by requiring an 
axino LSP with mass $m_{\ta}\sim 0.1-1$~MeV. Neutralinos produced 
either thermally or via gravitino decay will themselves decay 
typically to states such as $\ta\gamma$, so that the dark matter 
abundance is reduced by a factor $m_{\ta}/m_{\tz_1}$\cite{ckr}. 
The bulk of dark matter then resides in thermally produced axinos 
and/or in axions produced from vacuum mis-alignment. 
By driving up the value of PQ breaking scale $f_a/N$, thermal 
production of axinos is suppressed, and higher values of $T_R$ 
are required to maintain a total axino plus axion relic abundance  
of $\Omega_{a\ta}h^2\sim 0.11$.
To avoid overproduction of axions at high $f_a/N$, we adopted 
a small vacuum mis-alignment angle $\theta_i\sim 0.05$. However, 
the large values of $f_a/N\sim 10^{12}-10^{13}$~GeV suppress the 
$\tz_1$ decay rate, thus interfering with BBN from a different avenue. 
Models with a high-mass, bino-like $\tz_1$ and low 
``apparent'' $\Omega_{\tz_1}^{app}h^2$ can avoid the BBN bounds, 
and allow $T_R$ values in excess of $10^{10}$~GeV to be attained. 
As we showed, such conditions with $m_{\tG}\sim 10-30$~TeV can be reached
in Effective SUSY\cite{ckn,esusy} or mirage-unification 
SUSY breaking\cite{MU} models.

A related scenario to reconcile thermal leptogenesis with the gravitino 
problem --- using mixed axion/axino dark matter --- was proposed much 
earlier by Asaka and Yanagida (AY)\cite{ay}.
Their proposal was to work within the context of the PQMSSM, but with a sparticle 
mass hierarchy $m({\rm sparticle})>m_{\tG}>m_{\ta}$. In this case, the 
couplings of MSSM
sparticles to axinos are larger than the couplings to gravitinos, so that the
long-lived decays to gravitino are bypassed, and the sparticles are assumed to decay 
to an axino LSP shortly before the onset of BBN. Furthermore, thermally produced 
gravitinos decay inertly via $\tG\to a\ta$ and so do not disrupt BBN. Reheat temperatures
as high as $T_R\sim 10^{15}$ were claimed to be possible.

In this paper, we re-visit the AY scenario, incorporating several 
improvements into our analysis. In particular, we implement 
\begin{enumerate}
\item the latest astrophysically measured value of\cite{wmap7} 
\be
\Omega_{\rm DM}h^2= 0.1123\pm 0.0035\ \ \ {\rm at\ 68\%\ CL};
\ee
\item the latest calculations for thermal production of gravitinos and axinos; 
\item vacuum-misalignment production of axions as an element of the dark matter abundance; 
\item the latest BBN constraints on late decaying particles; and finally 
\item a careful treatment of entropy production from late decaying saxions.
Since entropy production from saxion decay will also dilute the matter-antimatter asymmetry
by a factor $r$ (to be defined later), in this case a re-heat temperature
$T_R\agt 2r\times 10^9$ GeV will be needed.
\end{enumerate}
The re-analysis of the AY scenario taking into account points 
1.--4.\ is presented in Sec.~\ref{sec:AY}, while entropy injection 
from saxion decay is discussed in detail in Sec.~\ref{sec:saxion}.
In Sec.~\ref{sec:conclude}, we present our final conclusions and 
consequences of the AY scenario for LHC physics and dark matter searches.

%========================================================================================
\section{Relic density of mixed axion/axino DM in the AY scenario}
\label{sec:AY}
%========================================================================================

\subsection{MSSM parameters}
\label{ssec:BMs}

The only relevant MSSM parameters for our analysis are the $\tz_1$ and $\tG$ masses 
$m_{\tz_1}$ and $m_{\tG}$, the $\tz_1$ bino component $v_4^{(1)}$ in the notation of \cite{wss}, 
and the $\tz_1$ abundance after freeze-out $\Omega_{\tz_1}$. 
The remaining of the MSSM parameters only impact the running of the gauge couplings and the
value of $\Omega_{\tG}$, which depend on all the gaugino masses (see Eq.~(\ref{eq:omegaG}) below).
However, as shown below, in the AY scenario with $T_R \gtrsim 10^{9}$~GeV,
the contribution from $\tG\to a\ta$ decay to the dark matter relic density is negligible.
Thus the dependence on the entire SUSY spectrum is very mild. 

For illustration we will use a generic mSUGRA scenario with $m_0=1000$~GeV, 
$m_{1/2}=1000$~GeV, $A_0=0$, $\tan\beta =55$ and $\mu >0$, which gives 
$m_{\tz_1}=430$~GeV and  $\Omega_{\tz_1}=0.04$, but treat $m_{\tz_1}$ and 
$\Omega_{\tz_1}$ as free parameters throughout most of the numerical analysis. 
The bino component of the $\tz_1$ wavefunction is important, since it determines the $\tz_1-\ta$ coupling. For
simplicity we will assume a purely bino neutralino, which is valid for a large portion of the mSUGRA parameter space.
We also take $m_{\tG} = m_{\tz_1}/2$ and $m_{\ta} < m_{\tG}$ in order to have an axino LSP with a gravitino NLSP.

\subsection{Mixed axion/axino abundance calculation}
\label{ssec:Oh2}

Here, we consider four mechanisms for dark matter production in the
AY scenario.

\bi
\item If the reheat temperature $T_R$ exceeds the axino decoupling temperature
\be
T_{dcp}=10^{11}\ {\rm GeV}\left(\frac{f_a/N}{10^{12}\ {\rm GeV}}\right)^2
\left(\frac{0.1}{\alpha_s}\right)^3 ,
\ee
axinos will be in thermal equilibrium,
with an abundance given by  
\be
\Omega_{\ta}^{TE}h^2 \simeq 0.38 \left(\frac{m_{\ta}}{1\ {\rm keV}}\right).
\ee
To avoid overproducing axino dark matter, the RTW bound~\cite{rtw}
then implies that $m_{\ta}<0.3$~keV.

In the case where $T_R<T_{dcp}$, 
the axinos are never in thermal equilibrium in the early 
universe.
However, they can still be produced thermally via radiation off of particles 
that are in thermal equilibrium~\cite{ckkr,steffen}. 
Here, we adopt a recent calculation of the thermally produced (TP)
axino abundance from Strumia~\cite{strumia}:
\be
\Omega_{\ta}^{\rm TP}h^2=1.24 g_3^4 F(g_3)\frac{m_{\ta}}{{\rm GeV}}
\frac{T_R}{10^4\ {\rm GeV}}\left(\frac{10^{11}}{f_a/N}\right)^2 ,
\ee
with $F(g_3)\sim 20 g_3^2\ln\frac{3}{g_3}$, and $g_3$ is the strong coupling 
constant evaluated at $Q=T_R$.
\item In supersymmetric scenarios with a quasi-stable neutralino, the
$\tz_1$s will be present in thermal equilibrium in the early universe, and will
freeze out when the expansion rate exceeds their interaction rate, 
at a temperature roughly $T_{fo}\sim m_{\tz_1}/20$.
The present day abundance can be evaluated by integrating the Boltzmann equation.
Several computer codes are available for this computation. Here we use the
code IsaReD~\cite{isared},
a part of the Isajet/Isatools package~\cite{isatools,isajet}.

In our case, each neutralino will undergo decay to the stable axino LSP, via decays
such as $\tz_1\to\ta\gamma$. 
Neutralinos may also decay via {\it e.g.} $\tz_1\to \tG\gamma$, but these
modes are suppressed by $1/m_{Pl}^2$ rather than $1/(f_a/N)^2$, and so the
decay to gravitinos is suppressed (one of the hallmarks of the AY scenario).
Thus, the non-thermally produced (NTP) axinos will
inherit the thermally produced neutralino number density, and we will 
simply have~\cite{ckkr}
\be
\Omega_{\ta}^{\tz}h^2=\frac{m_{\ta}}{m_{\tz_1}}\Omega_{\tz_1}^{TP}h^2  .
\ee
\item Since here we are attempting to generate reheat temperatures $T_R\agt 10^9$~GeV,
we must also include in our calculations the thermal production of gravitinos 
in the early universe.
We adopt the calculation of Pradler and Steffen in Ref.~\cite{relic_G}, who 
have estimated the thermal gravitino production abundance as
\be
\Omega_{\tG}^{\rm TP}h^2 =\sum_{i=1}^{3}\omega_ig_i^2(T_R)
\left(1+\frac{M_i^2(T_R)}{3m_{\tG}^2}\right)\ln\left(\frac{k_i}{g_i(T_R)}\right)
\left(\frac{m_{\tG}}{100\ {\rm GeV}}\right)\left(\frac{T_R}{10^{10}\ {\rm GeV}}
\right) \label{eq:omegaG},
\ee
where $\omega_i=(0.018,0.044,0.117)$, $k_i=(1.266,1.312,1.271)$, $g_i$ are the 
gauge couplings evaluated at $Q=T_R$ and $M_i$ are the gaugino masses also evaluated 
at $Q=T_R$.
For the temperatures we are interested in, this agrees within a factor of about 2 with 
the more recent calculation by Rychkov and Strumia~\cite{relic_G}, which is sufficient for 
our purposes.

Since the only kinematically allowed gravitino decay mode is to an axion plus an axino LSP, the
abundance of axinos from gravitino production is given by
\be
\Omega_{\ta}^{\tG}h^2=\frac{m_{\ta}}{m_{\tG}}\Omega_{\tG}^{TP}h^2 ,
\ee
while the abundance of axions from gravitino production is given by
\be
\Omega_a^{\tG}h^2=\frac{m_{a}}{m_{\tG}}\Omega_{\tG}^{TP}h^2 .
\ee

For axino masses in the MeV range and gravitino masses in the TeV range,
the prefactor above is extremely small, making the contribution from
gravitino decays to the axino relic abundance negligible, what allows us to evade overproduction of
dark matter via thermal gravitino production.
\item 
Here, we consider the scenario where the PQ symmetry breaks before the
end of inflation, so that a nearly uniform value of the axion field
$\theta_i\equiv a(x)/(f_a/N)$ is expected throughout the universe.
The axion field equation of motion implies that the axion field stays relatively
constant until temperatures approach the QCD scale $T_{QCD}\sim 1$~GeV.
At this point, the temperature-dependent axion mass term turns on, and a
potential is induced for the axion field.
The axion field rolls towards its minimum and oscillates,
filling the universe with low energy (cold) axions.
The expected axion relic density via this vacuum mis-alignment mechanism is
given by~\cite{vacmis}
\be
\Omega_a h^2\simeq 0.23 f(\theta_i)\theta_i^2
\left(\frac{f_a/N}{10^{12}\ {\rm GeV}}\right)^{7/6}
\label{eq:mis-alignment}
\ee
where $0< \theta_i<\pi$ and $f(\theta_i)$ is the so-called anharmonicity
factor. Visinelli and Gondolo~\cite{vacmis} parametrize the latter as
$f(\theta_i)=\left[\ln\left(\frac{e}{1-\theta_i^2/\pi^2}\right)\right]^{7/6}$.
The uncertainty in $\Omega_a h^2$ from vacuum mis-alignment is estimated
as plus-or-minus a factor of three.
\ei

In this paper, we will evaluate the mixed axion/axino relic density from 
the above four sources:
\be
\Omega_{a\ta}h^2=\Omega_{\ta}^{\rm TP}h^2+
\Omega_{\ta}^{\tz}h^2+\Omega_{\ta}^{\tG}h^2+\Omega_{a}^{\tG}h^2+\Omega_{a}h^2 .
\label{eq:ata}
\ee
Over much of parameter space, if $m_{\ta}$ is taken to be of order the MeV scale
or below, then the contributions $\Omega_{\ta}^{\tG}$, $\Omega_{\ta}^{\tz_1}$ and $\Omega_{a}^{\tG}$
are subdominant.

In Fig.~\ref{fig:Oh2_SUG2}, we illustrate in frame {\it a}) the relative importance
of the four individual contributions as a function of $f_a/N$,
for $\Omega_{\tz_1} h^2 = 10$, $m_{\tz_1} = 430$~GeV and  $m_{\tG}=0.5 m_{\tz_1}$.
For the axion/axino sector we take $\theta_i=0.05$ and $m_{\ta}=100$~keV.
The value of $T_R$ is adjusted such that $\Omega_{a\ta}h^2=0.1123$.
Low values of $\theta_i$ suppress axion production and allow higher values of
$f_a/N$ to be probed; the higher values of $f_a/N$ suppress thermal axino
production, thus allowing for higher $T_R$ values to compensate.

For low $f_a/N$ values, the TP axino contribution is dominant. But as $f_a/N$ increases,
the axion component grows. 
For higher values of $f_a/N$, the vacuum-misalignment produced axion component dominates,
and the dark matter is predominantly composed of cold axions.
The contribution of axino dark matter from $\tz_1$ and $\tG$ decays are always
negligible in this case.

In frame {\it b}) of Fig.~\ref{fig:Oh2_SUG2}, we show the value of $T_R$
which is needed to enforce the total abundance of mixed axion/axino dark matter
to be $\Omega_{a\ta}h^2=0.1123$. We show cases for $m_{\ta}=0.1$ and 1 MeV.
As $f_a/N$ increases, the axino-matter coupling decreases, and one would 
expect the thermal production of axinos to decrease. Since we enforce
$\Omega_{a\ta}h^2=0.1123$, then higher values of $T_R$ are needed to
compensate and enhance the thermal production of axinos\cite{axdm} (and gravitinos).
We see that the value of $T_R$ can be pushed to over $10^9$~GeV for
$m_{\ta}=1$ MeV, and to over $10^{10}$~GeV for $m_{\ta}=0.1$ MeV, thus allowing
high enough $T_R$ for thermal leptogenesis.

%%%%%%%%%%%%%%%%%%%%%%%%%%%%%%%%%%%%%%%%%%%%%%%%%%%%%%%%%%%%%%%%%%%
\FIGURE[t]{
\includegraphics[width=10cm]{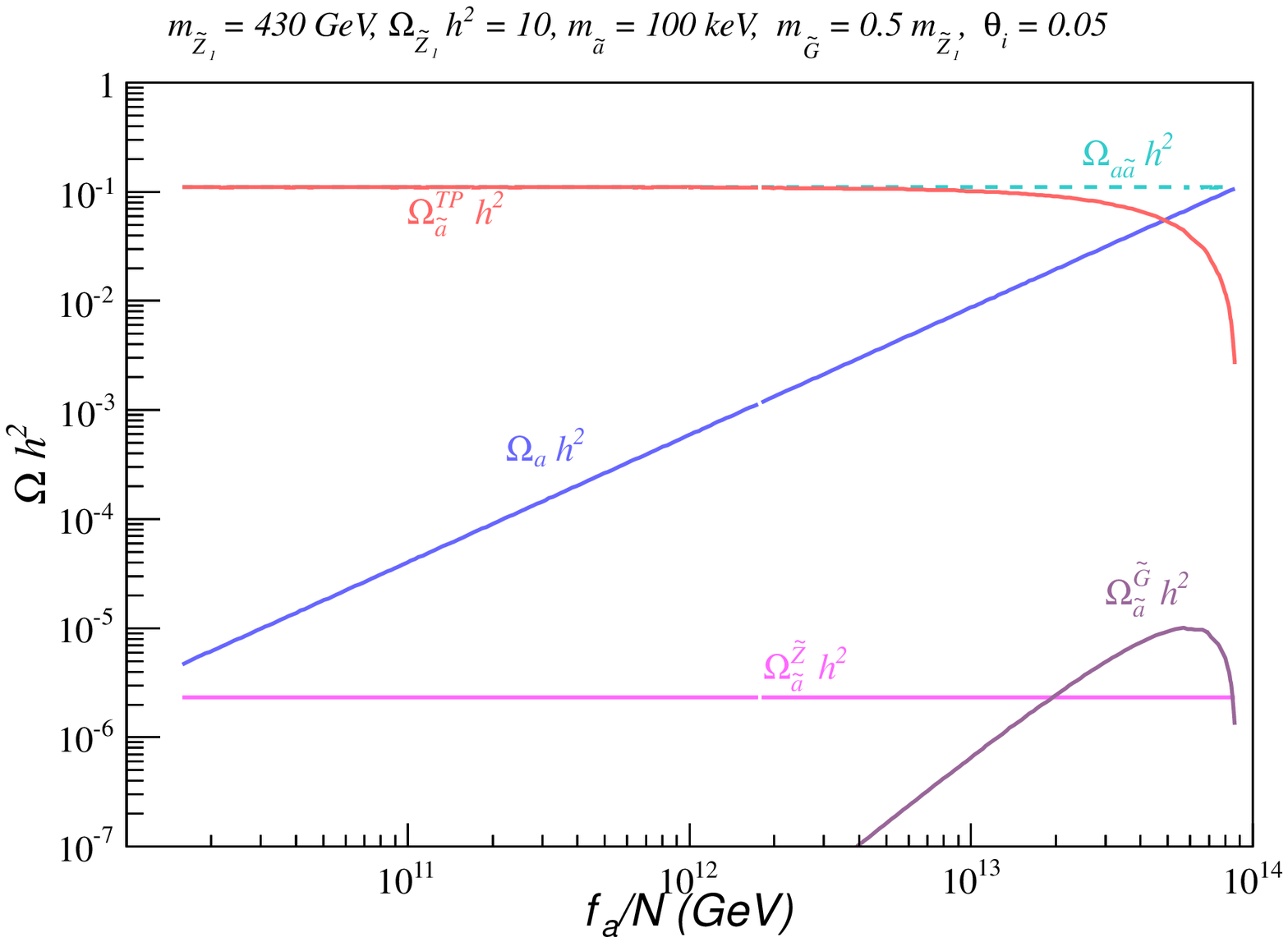}
\includegraphics[width=10cm]{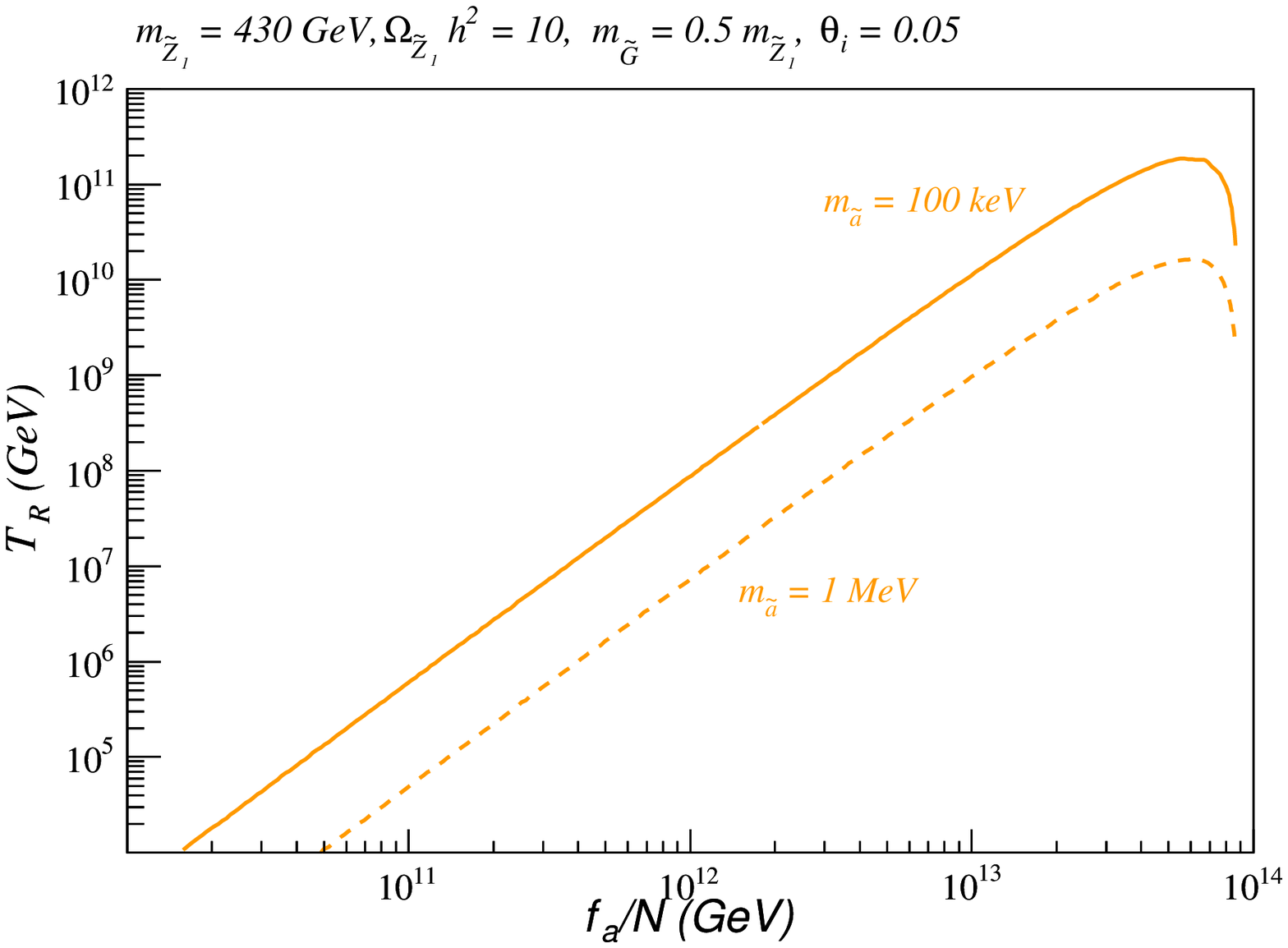}
\caption{Upper frame: Contribution of axions and TP and NTP  
axinos to the DM density as a function of the PQ breaking scale $f_a/N$, 
for $m_{\tz_1} = 430$~GeV, $\Omega_{\tz_1} h^2 = 10$,
 $m_{\ta}=100$~keV and $\theta_i=0.05$; $T_R$ is adjusted such that 
$\Omega_{a\ta}h^2=0.1123$. We assume $m_{\tG}=m_{\tz_1}/2$.
Lower frame: the value of $T_R$ that is needed to achieve $\Omega_{a\ta} h^2 = 0.1123$ 
for $m_{\ta} = 0.1$ and 1 MeV.
}\label{fig:Oh2_SUG2}}
%%%%%%%%%%%%%%%%%%%%%%%%%%%%%%%%%%%%%%%%%%%%%%%%%%%%%%%%%%%%%%%%%%%

\subsection{Constraints from cold/warm dark matter}
\label{ssec:cdm}

Depending on its mass, the axino might constitute 
warm (WDM) or hot (HDM) dark matter; the latter possibilities are severely constrained by the matter 
power spectrum and reionization~\cite{ckkr,jlm} (see also~\cite{warm,hot}).
We consider axinos with mass $1-100$~keV as mostly WDM, and axinos with mass $<1$~keV 
as mostly HDM. 
Since these bounds on the amount of WDM/HDM are model dependent~\cite{jlm}, 
we do not impose strict WDM/HDM constraints on our results. However, for guidance, we will keep
track of PQMSSM parameter points with potentially too large WDM and/or HDM components\footnote{Axions
produced from gravitino decay will also constitute HDM. However, since this contribution to the total
DM density is suppressed by $m_a/m_{\tG}$, it can be safely neglected.}:
As in \cite{Baer:2010kw}, we disfavour points with 
\be
  \begin{array}{lcl}
   \Omega_{\ta}/\Omega_{a\ta} > 0.2 &\forall& 1~{\rm keV} \le m_{\ta} < 100~{\rm keV},\ \ \ ({\rm WDM}) \\
   \Omega_{\ta}/\Omega_{a\ta} > 0.01 &\forall& m_{\ta} < 1~{\rm keV},\ \ \ ({\rm HDM})
  \end{array}
   \label{eq:whdm}
\ee
where $\Omega_{\ta} = \Omega_{\ta}^{TP} + \Omega_{\ta}^{\tG} + \Omega_{\ta}^{\tz}$. 
This is rather conservative. A rough estimate based on the neutrino mass 
limit~\cite{hot} from cosmological data, $\sum m_\nu<0.41$ to $0.44$~eV, 
gives that up to 4--5\% HDM contribution could be acceptable.
Moreover, 
Boyarsky {\it et al.} in \cite{warm} found that in case of a thermal relic (TR), 
100\% WDM is allowed for $m_{\rm TR}\ge 1.7$~keV, while for $m_{\rm TR}=1.1$~keV
as much as 40\% WDM is allowed at 95\% CL. We will also indicate these bounds, 
which are considerably weaker than Eq.~(\ref{eq:whdm}).

\subsection{Constraints on $\tz_1$ decay from BBN}
\label{ssec:bbn}

The AY scenario naturally avoids BBN constraints on late decaying gravitinos by
assuming the mass relation $m({\rm sparticle})>m_{\tG}>m_{\ta}$, so that the $\tG$
decays inertly 100\% of the time into $a\ta$. 
However, by searching for PQMSSM parameters which allow
$T_R\agt 2\times 10^9$~GeV while avoiding overproduction of mixed axino/axion dark matter
(the latter requires large $f_a/N\sim 10^{12}$~GeV and small $\theta_i$), 
we have pushed the $\tz_1$ lifetime uncomfortably high, so that its hadronic decays in the 
early universe now have the potential to disrupt BBN. The $\tz_1$ lifetime and hadronic 
branching fraction is calculated in Ref. \cite{ckkr,Baer:2010kw}.

Constraints from BBN on hadronic decays of long-lived neutral particles in the early universe
have been calculated in Ref's~\cite{ellis,kohri,jedamzik}. Here, we adopt the results
from Jedamzik~\cite{jedamzik}. The BBN constraints arise due to injection of 
high energy hadronic particles into the thermal plasma during or after BBN. The constraints depend
on three main factors:
\bi
\item The abundance of the long-lived neutral particles. In Ref.~\cite{jedamzik}, this
is given by $\Omega_X h^2$ where $X$ is the long-lived neutral particle which undergoes hadronic decays.
In our case, where the long-lived particle is the lightest neutralino which decays to an axino LSP,
this is just given by the usual thermal neutralino abundance $\Omega_{\tz_1}h^2$, as calculated
by IsaReD~\cite{isared}. 
\item The lifetime $\tau_X$ of the long-lived neutral particle. The longer-lived
$X$ is, the greater its potential to disrupt the successful BBN calculations. In our case
$\tau_X = \tau_{\tz_1} \propto (f_a/N)^2/m_{\tz_1}^3$.
\item The hadronic branching fraction $B_h$ of the long-lived neutral particle. If this is very small,  
then very little hadronic energy will be injected, and hence the constraints should be more mild.
\ei
From the above list, we see that BBN directly constrains the MSSM parameters ($\Omega_{\tz_1}h^2$
and $m_{\tz_1}$) of the PQMSSM model. The constraints also depend on $f_a$, since its value
directly affects $\tau_{\tz_1}$. The BBN constraints are shown in Fig.~9 (for $m_X=1$~TeV) and Fig.~10 (for $m_X=100$~GeV)
of Ref.~\cite{jedamzik}, as contours in the $\tau_X\ vs.\ \Omega_Xh^2$ plane, with numerous
contours for differing $B_h$ values ranging from $10^{-5}$ to 1. 
For $B_h\sim 0.1$, for instance, and very large values
of $\Omega_Xh^2\sim 10-10^3$, the lifetime $\tau_X$ must be $\alt 0.1$~sec, or else the
primordial abundance of $^4\rm He$ is disrupted. 
If $\Omega_X h^2$ drops below $\sim 1$, then much larger
values of $\tau_X$ up to $\sim 100$ sec are allowed. If one desires a long-lived hadronically
decaying particle in the early universe with $\tau_X\agt 100$ sec, then typically much lower
values of $\Omega_X h^2\sim 10^{-6}-10^{-4}$ are required. 

We have digitized the constraints of Ref.~\cite{jedamzik}, implementing
extrapolations for cases intermediate between values of parameters shown, so as to
approximately apply the BBN constraints to the AY scenario with a long-lived neutralino
decaying during BBN.
The results are shown in Fig.~\ref{fig:bbn}, as a function of $m_{\tz_1}$.
From Fig.~\ref{fig:Oh2_SUG2}{\it b)}, we have $T_R > 10^9$~GeV for $f_{a}/N \sim 10^{13}$~GeV.
From Fig.~\ref{fig:bbn}, we see that, for $m_{\tz_1} \sim 400$~GeV and $f_{a}/N = 10^{13}$~GeV,
we need $\Omega_{\tz_1}h^2 < 0.4$ in order to satisfy the BBN bounds. In particular, the assumed
value for $\Omega_{\tz_1}h^2$ (=10) in Fig.~\ref{fig:Oh2_SUG2} only satisfies the BBN constraints
for $f_a/N \lesssim 2\times 10^{12}$~GeV or $T_{R} \lesssim 3 \times 10^8$~GeV. Therefore we see that
the AY scenario with $T_R \gtrsim 10^9$~GeV requires quite low values of $\Omega_{\tz_1}h^2$ and/or
a heavy $\tz_1$.

%%%%%%%%%%%%%%%%%%%%%%%%%%%%%%%%%%%%%%%%%%%%%%%%%%%%%%%%%%%%%%%%%%%
\FIGURE[t]{
\includegraphics[width=10cm]{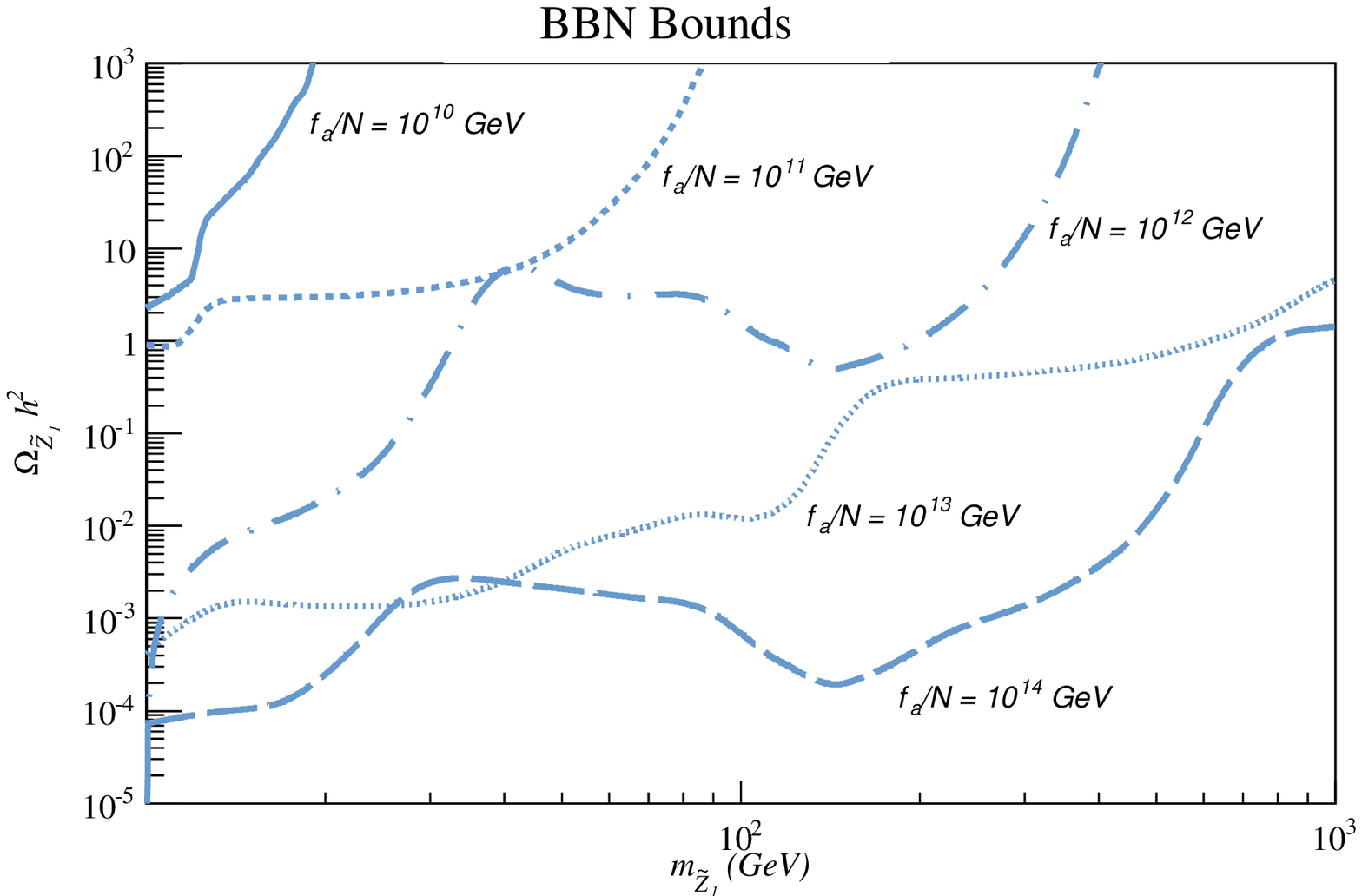}
\caption{BBN bounds on late decaying neutralinos ($\tz_1\to Z/\gamma + \ta$) from Ref. \cite{jedamzik} as
a function of $m_{\tz_1}$ for different values of $f_{a}/N$, assuming a bino $\tz_1$ and $m_{\tz_1} \gg m_{\ta}$.
The values of $\Omega_{\tz_1} h^2$ above the curves are excluded by BBN constraints.
}\label{fig:bbn}}
%%%%%%%%%%%%%%%%%%%%%%%%%%%%%%%%%%%%%%%%%%%%%%%%%%%%%%%%%%%%%%%%%%%

\subsection{Scan over PQ parameters}
\label{sec:scan}

The results of the last section were for a specific choice of the axino mass and
$\theta_i$ values. Next, we examine which values of $T_R$ are possible for arbitrary
values of $f_a$, $m_{\ta}$ and $\theta_i$. For now we keep the MSSM parameters 
($\Omega_{\tz_1}h^2$ and $m_{\tz_1}$) fixed. As discussed in the previous section,
the BBN constraints in general require low $\Omega_{\tz_1}h^2$ and high $m_{\tz_1}$. We
therefore assume $\Omega_{\tz_1}h^2 = 0.04$ and $m_{\tz_1} = 430$~GeV, which are
values consistent with, e.g., an mSUGRA point near the apex of the Higgs funnel region.
To probe the full PQ parameter space we perform a random scan over the PQ parameters in the range
\bea
 m_{\ta}  &\in& [10^{-7},\;10]\ {\rm GeV}\,,\nonumber \\
 f_a/N    &\in& [10^8,\;10^{15}]\ {\rm GeV}\,, \label{eq:scan}\\
 \theta_i &\in& [0,\;\pi]\,. \nonumber
\eea
and calculate the value of $T_R$ which is needed to enforce $\Omega_{a\ta}h^2=0.1123$.
The results of our scan are shown in Fig.~\ref{fig:sug2fa}, where we plot the derived 
value of $T_R$ versus PQ breaking scale $f_a/N$.
In the plot, dark blue and dark red points have mainly CDM with at most 20\%~WDM and 
at most 1\%~HDM admixture, c.f.\ Section~\ref{ssec:cdm}.
The light blue and light red points have higher values of WDM or HDM.
The red points are excluded by bounds derived from Ref. \cite{jedamzik}
on late decaying neutralinos which could destroy the succesful predictions of BBN.
Blue points are allowed by BBN constraints. 
Applying the WDM bounds of Boyarsky {\it et al.}\cite{warm} for a thermal relic 
and allowing up to 5\% HDM, the boundary between dark and light blue points 
moves left to the dashed blue line. 

We do see that a number
of dark blue points with mainly CDM, and which also respect BBN bounds, 
can be generated with $T_R>2\times 10^9$~GeV.  These 
thermal leptogenesis-consistent points all require
$f_a/N\agt 10^{12}$~GeV 
(or $f_a/N\agt 6\times 10^{11}$~GeV with weaker WDM/HDM requirements). 
%
%%%%%%%%%%%%%%%%%%%%%%%%%%%%%%%%%%%%%%%%%%%%%%%%%%%%%%%%%%%%%%%%%%%
\FIGURE[t]{
\includegraphics[width=10cm]{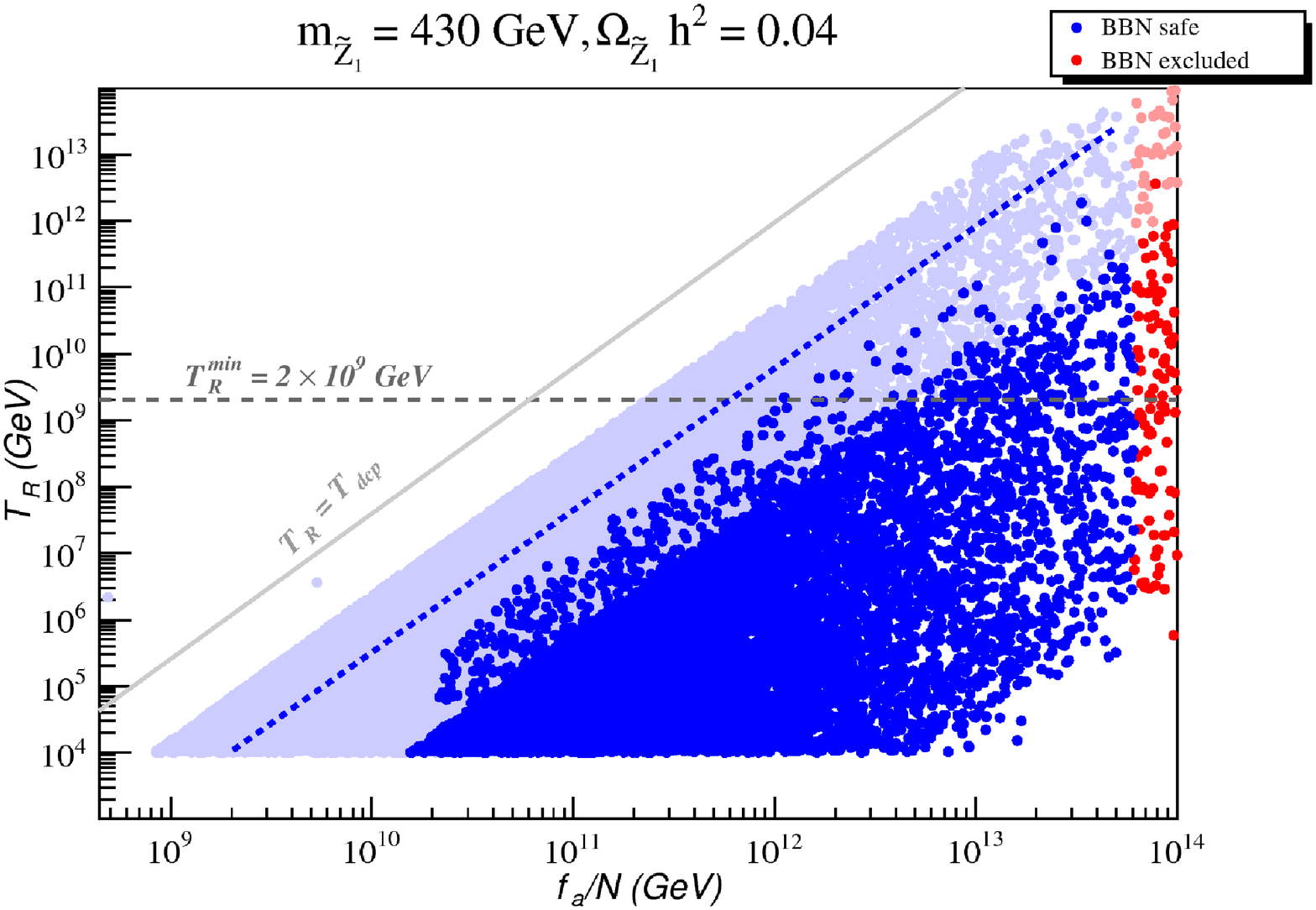}
\caption{Allowed and disallowed points in the $f_a\ vs.\ T_R$ plane
for $m_{\tz_1} = 430$~GeV and $\Omega_{\tz_1} h^{2} = 0.04$,
 including BBN constraints on late $\tz_1$ decay.
Dark blue points have mainly CDM with at most 20\%~WDM and 
at most 1\%~HDM admixture; the dashed blue line indicates 
the WDM limit by Boyarsky {\it et al.}\cite{warm} and up to 
5\% HDM}.
 \label{fig:sug2fa}}
%%%%%%%%%%%%%%%%%%%%%%%%%%%%%%%%%%%%%%%%%%%%%%%%%%%%%%%%%%%%%%%%%%%

In Fig.~\ref{fig:sug2th}, we show the same scan in the $\theta_i\ vs.\ T_R$
plane. Here, we see that the CDM/BBN consistent points with
high $T_R$ all need rather small values of axion mis-alignment angle
$\theta_i\alt 0.5$. This is needed since, at large $T_R$, large $f_a/N$ is 
necessary to suppress overproduction of axinos, while 
small $\theta_i$ is needed to suppress over-production of axions.
%
%%%%%%%%%%%%%%%%%%%%%%%%%%%%%%%%%%%%%%%%%%%%%%%%%%%%%%%%%%%%%%%%%%%
\FIGURE[t]{
\includegraphics[width=10cm]{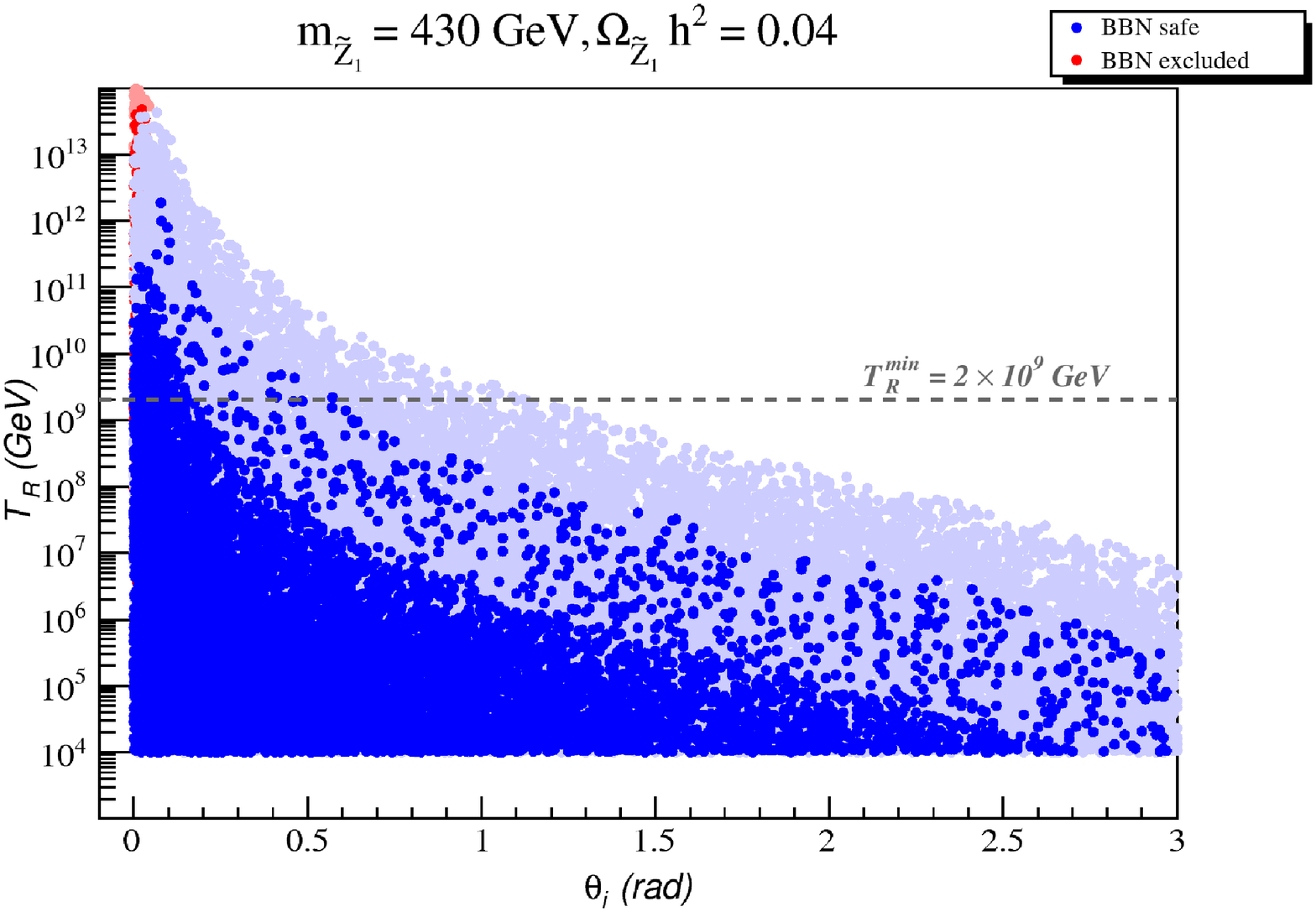}
\caption{Allowed and disallowed points in the $\theta_i\ vs.\ T_R$ plane
for $m_{\tz_1} = 430$~GeV and $\Omega_{\tz_1} h^{2} = 0.04$, 
including BBN constraints on late $\tz_1$ decay.
}\label{fig:sug2th}}
%%%%%%%%%%%%%%%%%%%%%%%%%%%%%%%%%%%%%%%%%%%%%%%%%%%%%%%%%%%%%%%%%%%

In Fig.~\ref{fig:sug2axion}, we show the same PQMSSM parameter scan for
$m_{\tz_1} = 430$~GeV and $\Omega_{\tz_1} h^{2} = 0.04$, but in the $\Omega_a h^2\ vs.\ T_R$
plane. Here, we see that the CDM/BBN consistent points with
high $T_R$ can have both large and small values of $\Omega_a h^2$.
Solutions with $\Omega_a h^2 \sim 0.1$ usually have very light axinos
(to suppress $\Omega_{\ta} h^2$) and moderate $\theta_i$ values. 
Solutions with $\Omega_a h^2 \lesssim 0.1$ usually have heavier axinos
and small $\theta_i$.
%
%%%%%%%%%%%%%%%%%%%%%%%%%%%%%%%%%%%%%%%%%%%%%%%%%%%%%%%%%%%%%%%%%%%
\FIGURE[t]{
\includegraphics[width=10cm]{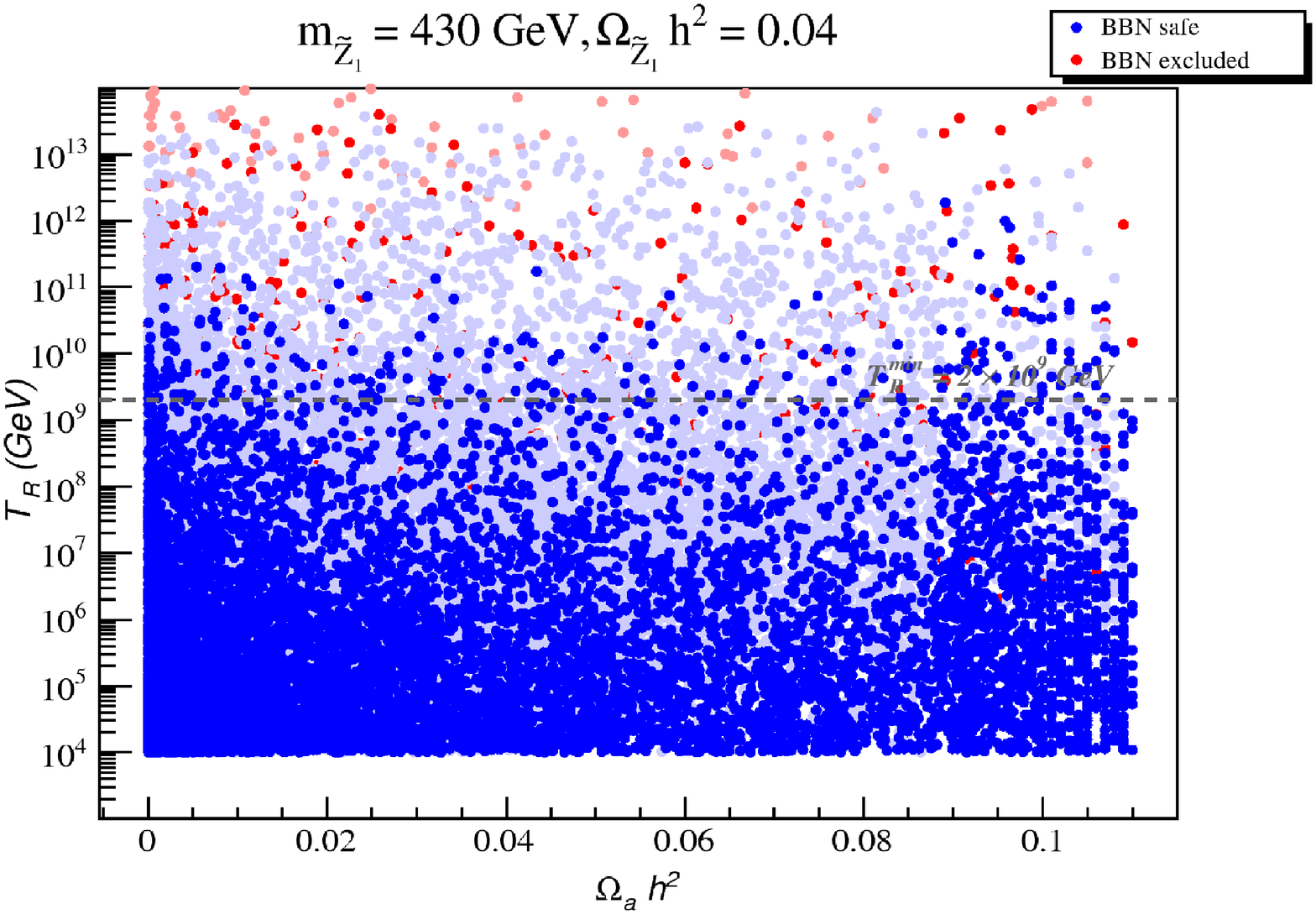}
\caption{Allowed and disallowed points in the $\Omega_a h^2\ vs.\ T_R$ plane
for $m_{\tz_1} = 430$~GeV and $\Omega_{\tz_1} h^{2} = 0.04$,
 including BBN constraints on late $\tz_1$ decay.
}\label{fig:sug2axion}}
%%%%%%%%%%%%%%%%%%%%%%%%%%%%%%%%%%%%%%%%%%%%%%%%%%%%%%%%%%%%%%%%%%%

While the results of the previous figures are restricted to specific
values of $m_{\tz_1}$ and $\Omega_{\tz_1} h^{2}$, the overall scheme is much more general.
The lesson here is that the AY scenario for reconciling thermal leptogenesis with the gravitino
problem can work provided certain conditions on SUSY models are met. These conditions
are rather similar to those needed by the large $m_{3/2}$ scenario put forth
in Ref. \cite{Baer:2010kw}. After adopting a model with a sparticle mass hierarchy of
$m({\rm sparticles})>m_{\tG}>m_{\ta}$, with $m_{\ta}\sim $MeV scale and 
$m_{\tG}\sim M_{{\rm weak}}$, one needs the following features:
\bi 
\item To allow for $T_R>2\times 10^9$~GeV, one must suppress thermal production of axinos via a large
value of $f_a/N\agt 10^{12}$~GeV.
\item To suppress overproduction of axions one must adopt a lower range of mis-alignment 
angle $\theta_i\alt 0.5$ (or $\theta_i\alt 0.8$ taking into account the factor 3 uncertainty 
in Eq.~(\ref{eq:mis-alignment})) .
\item The large value of $f_a/N$ increases the $\tz_1$ lifetime, which brings in BBN
constraints on late-decaying neutral particles. To avoid BBN bounds, it helps to invoke
1.~a~bino-like $\tz_1$ so that $v_4^{(1)}\sim 1$, 2.~a~low apparent neutralino 
relic abundance $\Omega_{\tz_1}^{app} h^2\alt 1$ and 3.~a~large value of $m_{\tz_1}$ to
help suppress the $\tz_1$ lifetime.
\ei

\noindent
These conditions are illustrated in a more model independent way in Fig.~\ref{fig:OmMZ1}. 
Here, we assume the AY mass hierarchy, but extend 
our previous scan to the whole PQMSSM parameter space:
\bea
 m_{\ta}  &\in& [10^{-7},\;10]\ {\rm GeV}\,,\nonumber \\
 f_a/N    &\in& [10^8,\;10^{15}]\ {\rm GeV}\,, \nonumber\\
 \theta_i &\in& [0,\;\pi]\,,\nonumber\\
  \Omega_{\tz_1} h^2 &\in& [10^{-5},10^{3}]\,,\nonumber\\
  m_{\tz_1} &\in& [10 ,10^4]\ {\rm GeV}\,. \nonumber
\eea
As before, we assume $m_{\tG} = m_{\tz_1}/2$ and the blue points are BBN-allowed, while red points violate BBN bounds.
The dashed line indicates the boundary blow which 99\% of the DM/BBN consistent 
solutions lie when applying weaker WDM/HDM requirements (WDM limit according to 
Boyarsky {\it et al.}\cite{warm} and up to 5\% HDM, cf.\ Fig.~\ref{fig:sug2fa}) 
This line can be interpreted as a
natural upper bound for $\Omega_{\tz_1} h^2$ as a function of $m_{\tz_1}$.
From this we see that models with $m_{\tz_1} \lesssim 10$~GeV require 
$\Omega_{\tz_1} h^2\lesssim 10^{-3}$, while values of $\Omega_{\tz_1} h^2$ as 
high as $10^{3}$ can be consistent with thermal leptogenesis if the neutralino 
is in the TeV range.

%%%%%%%%%%%%%%%%%%%%%%%%%%%%%%%%%%%%%%%%%%%%%%%%%%%%%%%%%%%%%%%%%%%
\FIGURE[t]{
\includegraphics[width=10cm]{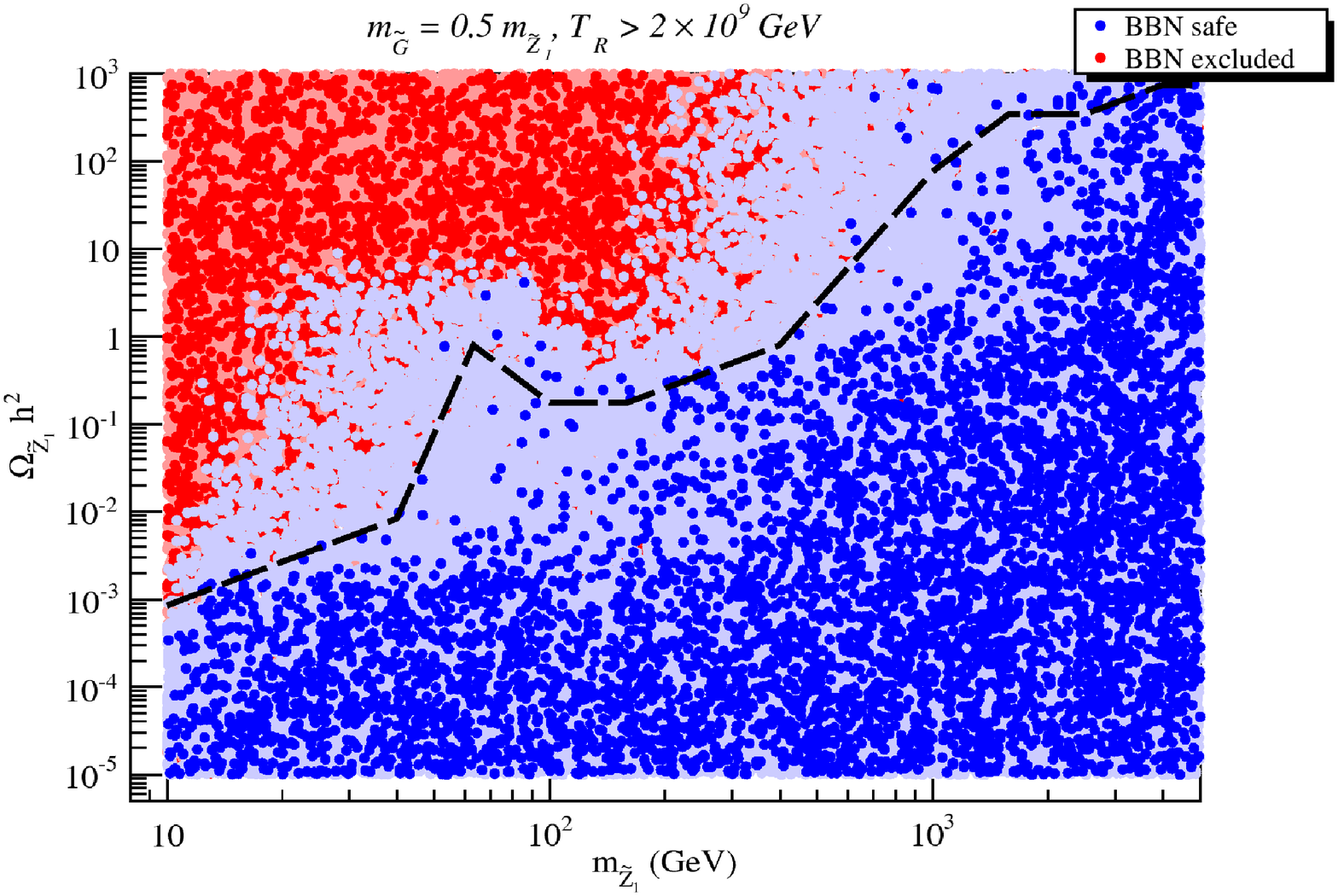}
\caption{Allowed and disallowed points in the $\Omega_{\tz_1} h^2\ vs. m_{\tz_1}$ plane
for a general scan over SUSY models with a bino $\tz_1$.
For all points, we require $T_R>2\times 10^9$~GeV and assume $m_{\tG} = m_{\tz_1}/2$.
Dark blue points are consistent with BBN and have mainly CDM with at 
most 20\%~WDM and/or 1\%~HDM admixture.  
The region below the dashed line represents the MSSM parameter space
where 99\% of the DM/BBN consistent solutions lie when applying weaker 
WDM/HDM requirements as discussed in the text.
}\label{fig:OmMZ1}}
%%%%%%%%%%%%%%%%%%%%%%%%%%%%%%%%%%%%%%%%%%%%%%%%%%%%%%%%%%%%%%%%%%%

%========================================================================================
\section{Dilution of DM by entropy production from saxion decay}
\label{sec:saxion}
%========================================================================================

Up to this point, we have neglected an important element of the axion supermultiplet, 
namely the spin-0 saxion field $s(x)$ which is expected to obtain a soft SUSY breaking mass
at the GUT scale of $m_s\sim m_0$. In the same way as axinos, saxions can be produced in
thermal equilibrium (if $T_R > T_{dcp}$) or out of equilibrium
 from scatterings of particles in the plasma (if $T_R < T_{dcp}$).
However a second mechanism of saxion production is also possible \cite{kns}. 
After supersymmetry breaking, the saxion potential $V(s)$ develops
a global minimum, causing the saxion field to coherently oscillate around its minimum. This
coherent oscillation can have a large energy density, which contributes to the total
saxion energy density if $T_R < T_{dcp}$. However, if $T_R > T_{dcp}$, the coherent
oscillating saxions will couple to the thermal plasma and thermalize.

Once the saxions decouple from the thermal plasma (at $T=T_{dcp}$) and become non-relativistic
(at $T\sim m_s$), their energy density ($\rho_s$) scales as $T^3$ (or $R^{-3}$), 
while the thermal plasma's energy density ($\rho_{rad}$) scales as $T^4$ (or $R^{-4}$). 
If the saxion lifetime is sufficiently long, at some temperature $T_e$, 
we will have $\rho_{s} > \rho_{rad}$ and the universe will become temporarily matter dominated until the saxions decay.

Being a {\it R}-parity even state, the saxion can decay to standard model states or pairs
of sparticles. Since we assume $m_s\sim m_0$, the decay into SUSY states will be 
kinematically suppressed and the saxion decays will mostly consist of SM particles.
Therefore the saxion decay products will thermalize in the thermal plasma, which 
is then ``relatively reheated"\cite{turner} with respect to other decoupled particles, 
such as axinos. As a consequence, all particles decoupled
from the thermal plasma during the saxion decay will have their number density diluted with respect
to the thermal bath's. 
Below we introduce the relevant expressions necessary for computing this
dilution factor ($r$) and in Sec. \ref{sec:rscan} we discuss how the inclusion of the saxion field
impacts our previous results. 
Since entropy injection from saxion decay will also dilute the matter-antimatter asymmetry
by a factor $r$, 
in this case a re-heat temperature $T_R\agt 2r\times 10^9$ GeV will be required for a successful implementation of thermal leptogenesis.

\subsection{Saxion production and decay}
\label{ssec:sprod}

As mentioned above, if $T_R$ exceeds $T_{dcp}$ in the early universe, saxions
are produced in thermal equilibrium such that
\be
Y_s m_s = \frac{\rho_s}{s} \simeq 10^{-3} \frac{m_s}{\rm{GeV}},
\ee
where $s=2 \pi^2 g_{*} T^3/45$ is the plasma entropy density and $Y_s$ is the saxion yield.
For $T_R<T_{dcp}$, saxions can still be produced thermally,
although to our knowledge a full calculation is not yet available.
In Ref's \cite{ay} and \cite{kns}, thermal saxion production is
estimated to be
\be
\frac{\rho_s}{s}\simeq 10^{-3} m_s T_R/T_{dcp}=m_s \left(\frac{T_R}{10^{14}\ {\rm GeV}}\right)
\left(\frac{10^{12}\ {\rm GeV}}{f_a/N}\right)^2 ,
\ee
which we will adopt for our calculations.

In addition, saxions can be produced via coherent oscillations of the
saxion field in the early universe. Although the exact mechanism
depends on the saxion potential near the SUSY breaking scale, the energy
density associated with the coherent oscillations can be parametrized
by the initial saxion field strength, $s_i$. Natural values for $s_i$
are $s_i \sim f_a$ or $s_i \sim M_{Pl}$. Unless stated otherwise,
we will assume $s_i = f_a/N$. The saxion energy density is estimated
for the case of very high values of $T_R$ with 
$\Gamma_I> m_s$ (here, $T_R$ is related to the inflaton decay width
$\Gamma_I$ as $T_R=(3/g_*\pi^3)^{1/4}(M_{Pl}\Gamma_I)^{1/2}$)
as \cite{kns}
\be
\frac{\rho_s}{s}\simeq 1.5\times 10^{-5}\ {\rm GeV}\left(\frac{m_s}{1\ {\rm GeV}}\right)^{1/2}
\left(\frac{(f_a/N)}{10^{12}\ {\rm GeV}}\right)^2\left(\frac{s_i}{(f_a/N)}\right)^2
\ee
while for $\Gamma_I<m_s$,
\be
\frac{\rho_s}{s}\simeq 2.1\times 10^{-9}\ {\rm GeV}\left(\frac{T_R}{10^5\ {\rm GeV}}\right)
\left(\frac{(f_a/N)}{10^{12}\ {\rm GeV}}\right)^2\left(\frac{s_i}{(f_a/N)}\right)^2 .
\ee
The summed saxion abundance is then given by the thermal production if $T_R > T_{dcp}$
or by the sum of thermal production plus the abundance from coherent oscillations, if $T_R < T_{dcp}$.

As an example, we show in Fig.~\ref{fig:SaxYield} the saxion yield $Y_s$ versus 
$f_a/N$ for $m_s =0.1$ and 1 TeV and for $T_R=10^9$~GeV. At low $f_a/N$,
$T_R>T_{dcp}$, and saxions are produced in thermal equilibrium. Once
$T_{dcp}$ rises above $T_R$, thermal saxion production dominates, 
but decreases with increasing $f_a/N$ until the point where saxion production from
coherent oscilations dominates.
%
%%%%%%%%%%%%%%%%%%%%%%%%%%%%%%%%%%%%%%%%%%%%%%%%%%%%%%%%%%%%%%%%%%%
\FIGURE[t]{
\includegraphics[width=10cm]{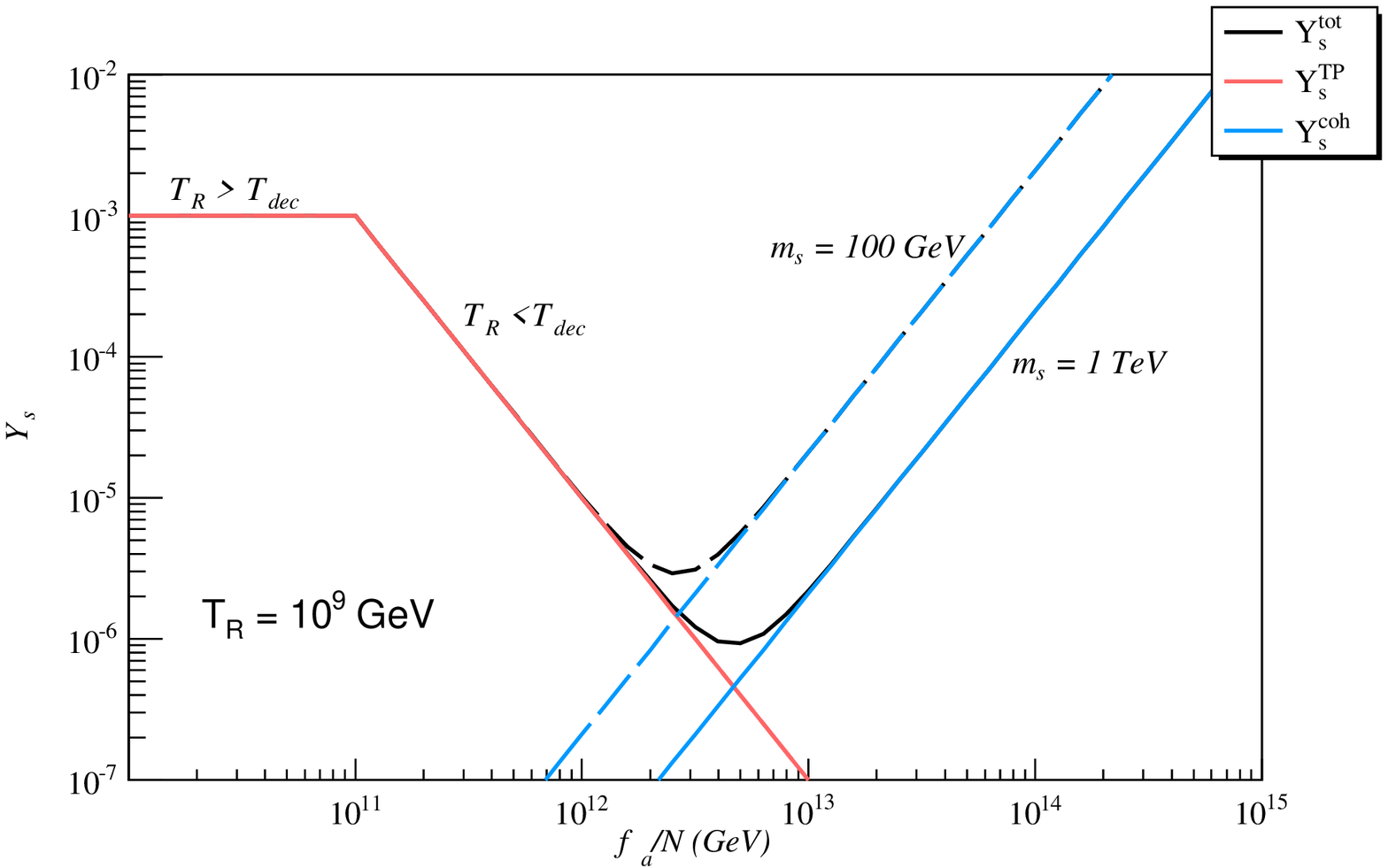}
\caption{Saxion yield $Y_s$ versus $f_a/N$ for $T_R=10^9$~GeV
and $m_s=0.1$ and 1 TeV. We assume $s_i/(f_a/N)=1$.
}\label{fig:SaxYield}}
%%%%%%%%%%%%%%%%%%%%%%%%%%%%%%%%%%%%%%%%%%%%%%%%%%%%%%%%%%%%%%%%%%%

The saxion is an $R$-parity even state which is
expected to dominantly decay into two gluons: $s\to gg$.
The saxion decay width differs by factors of two in Ref's \cite{kim}, \cite{ay} 
and \cite{kns}.
By an independent calculation, we find 
\be
\Gamma (s\to gg)=\frac{\alpha_s^2m_s^3}{32\pi^3 (f_a/N)^2},
\ee
in agreement with \cite{ay}.
The saxion may also decay (or not, model dependently) 
via $s\to aa$, and in the DFSZ\cite{dfsz} model, into 
$q\bar{q}$ or $\ell\bar{\ell}$. These latter decays are suppressed in the
KSVZ model\cite{ksvz}. Saxion may also decay to $\tz_i\tz_j$, $\gamma\gamma$ and $\tg\tg$. 
For saxion decay to gluino pairs, we calculate the interaction as\footnote{Our
saxion-gluino-gluino interaction differs by a factor of 2 from Ref. \cite{strumia}.}
\be
{\cal L}\ni \frac{\alpha_s}{4\pi (f_a/N)} s{\bar{\tg}}_A (i \Dsl )_{AB}\tg_B
\ee
and then find
\be
\Gamma (s\to \tg\tg )=\frac{\alpha_s^2m_s m_{\tg}^2}{8\pi^3 (f_a/N)^2}
\left(1-\frac{4m_{\tg}^2}{m_s^2}\right)^{3/2} .
\ee
The two widths are compared in Fig.~\ref{fig:Gam_s}, where the $s\to gg$ decay 
is found to always dominate. 
%
%%%%%%%%%%%%%%%%%%%%%%%%%%%%%%%%%%%%%%%%%%%%%%%%%%%%%%%%%%%%%%%%%%%
\FIGURE[t]{
\includegraphics[width=10cm]{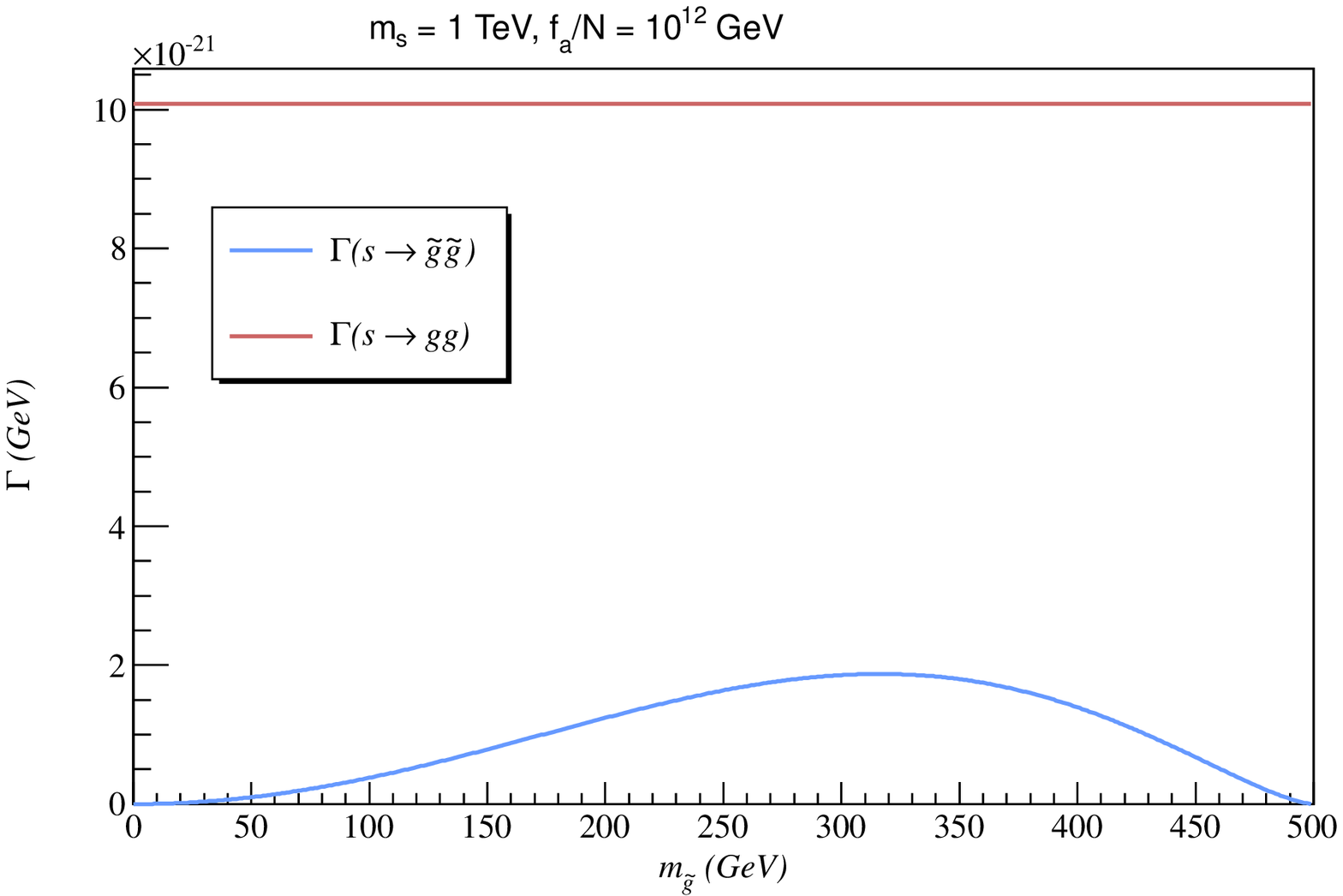}
\caption{Decay widths for $s\to gg$ and $s\to\tg\tg$ as a function
of $m_{\tg}$ for $f_a/N=10^{12}$ GeV and $m_s=1$~TeV.
}\label{fig:Gam_s}}
%%%%%%%%%%%%%%%%%%%%%%%%%%%%%%%%%%%%%%%%%%%%%%%%%%%%%%%%%%%%%%%%%%%

The temperature associated with saxion decay and entropy
injection is given by \cite{turner}
\be
%T_s=\sqrt{\Gamma_s M_{Pl}}/(\pi^2 g_*/90)^{1/4} .
T_s\simeq 0.78 g_*^{-1/4}\sqrt{\Gamma_s M_{Pl}} .
\ee
All the saxion decays essentially finish entropy injection by the time 
the universe cools to this value\cite{turner}.
For simplicity, here we will assume the $\Gamma_s=\Gamma (s\to gg)$ so that our results
are independent of $m_{\tg}$. Folding in the additional strong decay $s\to \tg\tg$
will result typically in a small increase in $T_s$.
In Fig.~\ref{fig:Tsax}, we plot the value $T_s$ as a function of $f_a/N$ for
three different values of the saxion mass.
%
%%%%%%%%%%%%%%%%%%%%%%%%%%%%%%%%%%%%%%%%%%%%%%%%%%%%%%%%%%%%%%%%%%%
\FIGURE[t]{
\includegraphics[width=10cm]{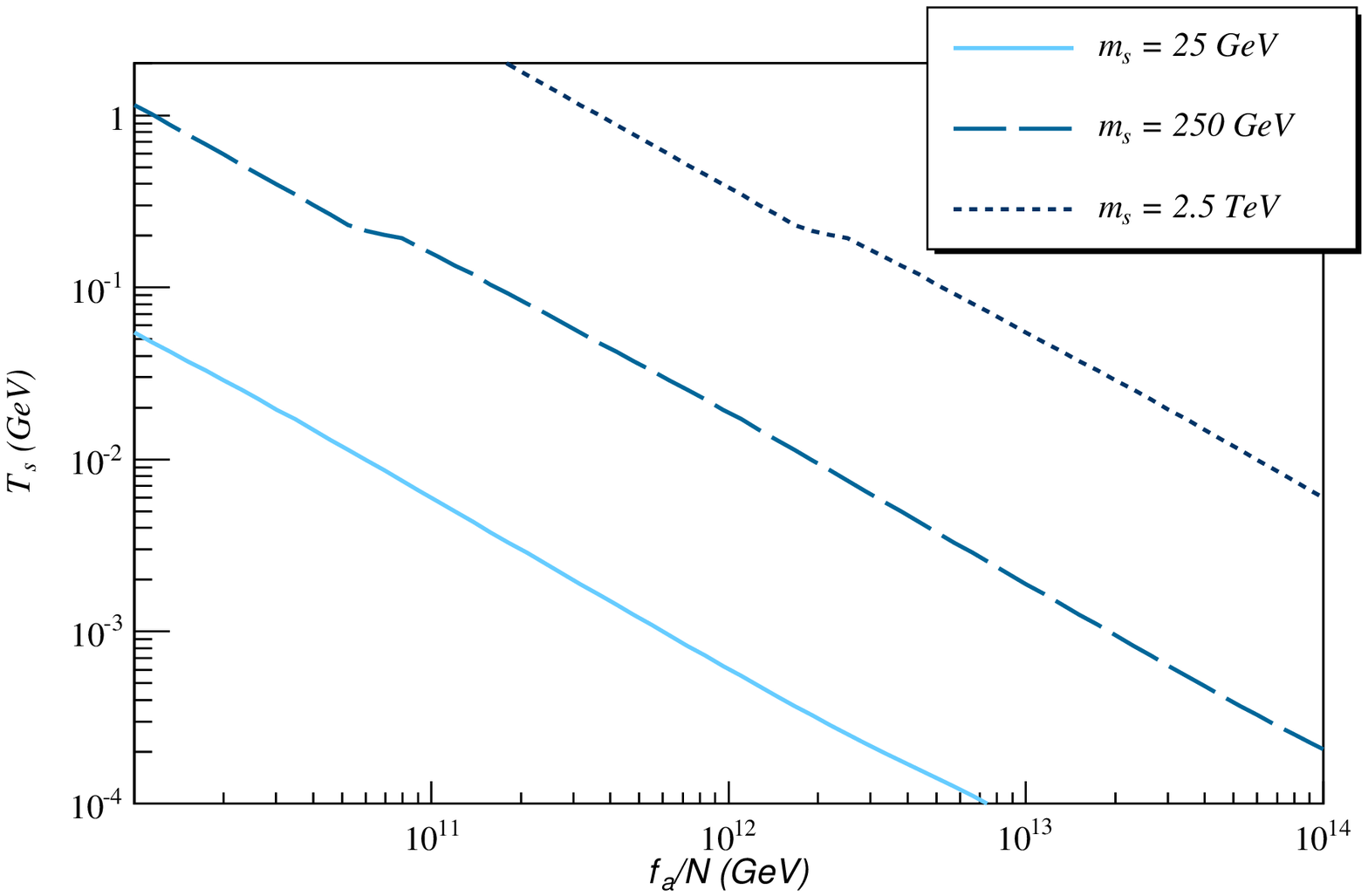}
\caption{Temperature $T_{s}$ at which saxions finish injecting entropy
versus $f_a/N$ for $m_s=25,\ 250\ {\rm and}\ 2500$~GeV.
}\label{fig:Tsax}}
%%%%%%%%%%%%%%%%%%%%%%%%%%%%%%%%%%%%%%%%%%%%%%%%%%%%%%%%%%%%%%%%%%%

\subsection{Entropy injection from saxion decay}

Armed with expressions for the saxion production rate in the early universe, we
next calculate the temperature $T_e$ at which the saxion energy density in the universe
equals the plasma energy density:
\be
\rho_s(T_e) = \rho_{rad}(T_e)=\frac{\pi^2g_*}{30} T_e^4.
\ee
Using $\rho_s = m_s Y_s s$ and $s=\frac{2\pi^2}{45}g_{*}T^3$, 
we find 
\be
T_e={4\over 3}m_s Y_s .
\ee
If $T_e$ exceeds $T_s$ ({\it i.e.} if saxion domination occurs before saxion decay),
then saxions dominate the energy density of the universe for $T_s \lesssim T \lesssim T_e$. 
In this case, saxion decays may inject significant entropy
and dilute whatever abundances are present at temperature $T_s$.
The situation is shown in Fig.~\ref{fig:Ts}, where we show the value of $T_s$ (blue horizontal lines)
and the value of $T_e$ (red lines) for $f_a/N=10^{12}$ and $10^{14}$~GeV. For
$T_R$ greater than the $r=1$ intersection points, significant entropy injection can occur.
%
%%%%%%%%%%%%%%%%%%%%%%%%%%%%%%%%%%%%%%%%%%%%%%%%%%%%%%%%%%%%%%%%%%%
\FIGURE[t]{
\includegraphics[width=10cm]{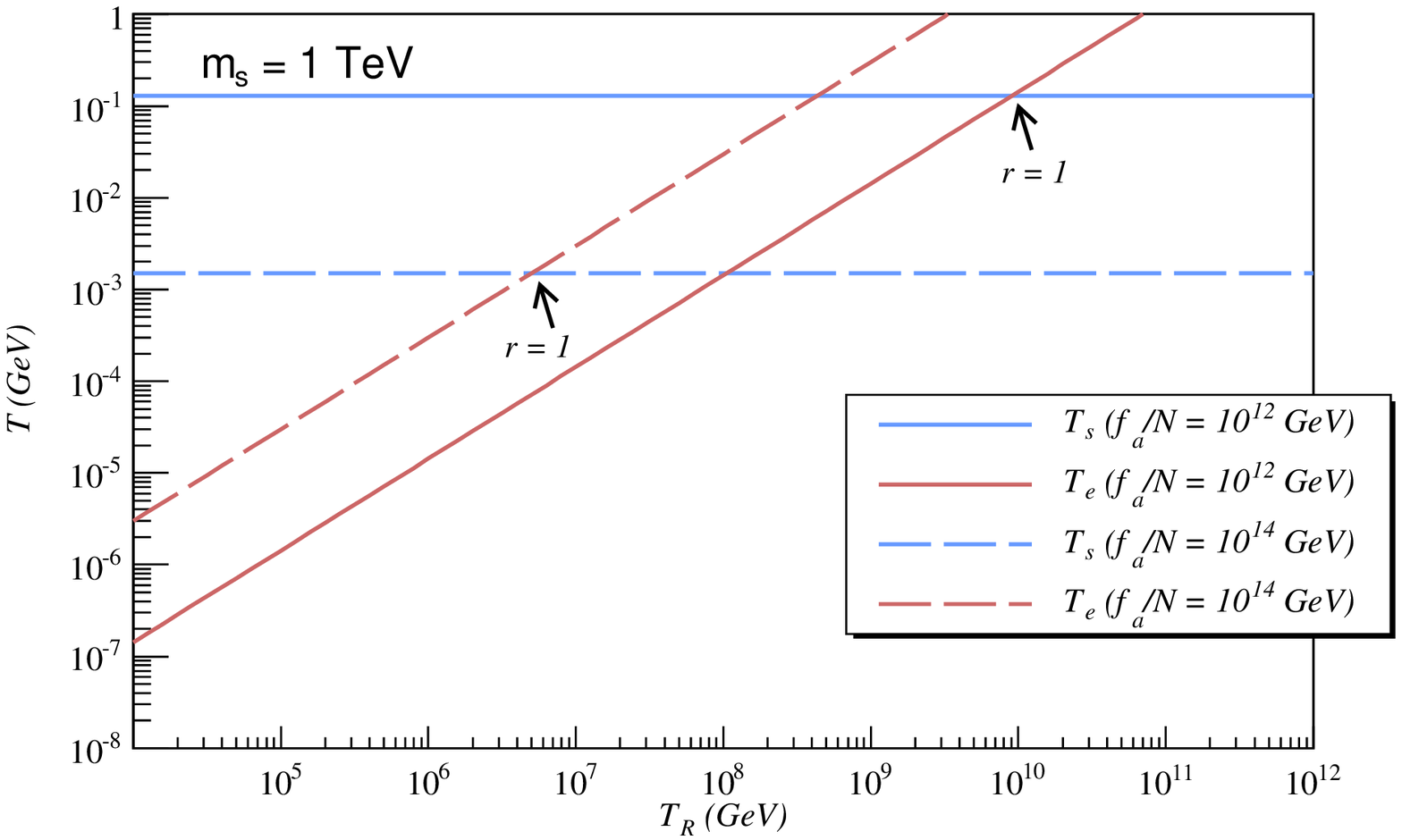}
\caption{Temperatures $T_{s}$ and $T_e$ versus $T_R$ for
$f_a/N=10^{12}$ and $10^{14}$~GeV and for $m_s=1$~TeV.
}\label{fig:Ts}}
%%%%%%%%%%%%%%%%%%%%%%%%%%%%%%%%%%%%%%%%%%%%%%%%%%%%%%%%%%%%%%%%%%%

If $r>1$, $T_s$ must satisfy $T_s\agt 10$ MeV, so the universe becomes radiation dominated before
BBN starts. On the other hand, if $r<1$, the early universe is always radiation dominated
and the usual BBN constraints on late decaying particles can be applied\cite{jedamzik}, as in
the case of neutralino decays.
%Furthermore, if the saxion decay temperature exceeds $T_{fo}$, the temperature
%at which neutralinos freeze out ($T_{fo}\sim m_{\tz_1}/20$), their decay products will just 
%be thermalized, and not contribute to dilution of dark matter.
If the conditions $10\ {\rm MeV} <T_s$ and $T_e>T_s$ hold, then saxion decay can inject substantial
entropy and dilute whatever relics are present and decoupled from the thermal plasma at the time
of saxion decay ($T=T_s$).
The ratio of entropy injection before and after a quasi-stable particle decay, for a matter dominated
universe ($T_e > T_s$), has been
calculated in Scherrer and Turner\cite{turner}, and is given by
\be
r=\frac{S_f}{S_i}\simeq 1.83\bar{g}_*^{1/4}\frac{Y_s m_s}{(M_{Pl}\Gamma_s)^{1/2}} \sim \frac{T_e}{T_s} ,
\ee
where $\bar{g}_*$ is the number of relativistic degrees of freedom averaged over the saxion decay
period, which we approximate by $g_{*}(T_s)$. The above expression for $r$ is only valid for $T_e > T_s$
(saxion dominated universe) or $r>1$. However, if the saxion energy density never dominates the universe,
the entropy injection is negligible\cite{turner}. Therefore we assume $r=1$ (no entropy injection), if $T_e < T_s$. 

Assuming $\Gamma_s$ is dominated by the $s\to gg$ decay, we  plot in Fig.~\ref{fig:R}{\it a})
the value of $r$ in the $(f_a/N)\ vs.\ T_R$ plane for $m_s=0.1,\ 1\ {\rm and}\ 10$~TeV, 
assuming $s_i/(f_a/N)=1$ (for production from coherent oscillations). The solid lines
all maintain $T_s>10$ MeV, while dashed lines violate this constraint. We see first that for
$m_s=100$~GeV, significant entropy production only occurs for $f_a/N\alt 4\times 10^{11}$~GeV; 
for larger $f_a/N$, $\Gamma_s$ is suppressed and the saxion lives long enough to decay
during or after BBN. For $m_s=1$~TeV, entropy injection can occur for
$f_a/N\alt 10^{13}$~GeV.

If $T_R$ lies below the $r=1$ contours, then not much entropy is injected, but for
high $T_R$, large entropy injection is possible and must be accounted for.
The various contours of constant $r$ initially increase with $T_R$. In this case,
the saxion production is dominantly thermal. When the curves turn over,  
saxion production is dominated by coherent oscillations. In this case, as
$f_a/N$ increases, the saxion field strength also increases (since $s_i/(f_a/N)$ is fixed to 1), 
and much lower $T_R$ values are allowed for substantial entropy production. 
Another noteworthy feature is that
the contours of entropy production increase with $T_R$ as $m_s$ increases. 
Thus, the dilution of DM from saxion decay can be reduced by requiring rather heavy saxions.
Finally, when we compare Fig.~\ref{fig:sug2fa} to Fig.~\ref{fig:R}, we see that
the range of $T_R\sim 10^9-10^{11}$~GeV for $f_a/N\sim 10^{12}-10^{14}$ implies that
entropy dilution from saxion decay needs to be accounted for in our calculations for
the case where $s_i/(f_a/N)\sim 1$ and $m_s = m_0 = 1$~TeV.
%
%%%%%%%%%%%%%%%%%%%%%%%%%%%%%%%%%%%%%%%%%%%%%%%%%%%%%%%%%%%%%%%%%%%
\FIGURE[t]{
\includegraphics[width=11cm]{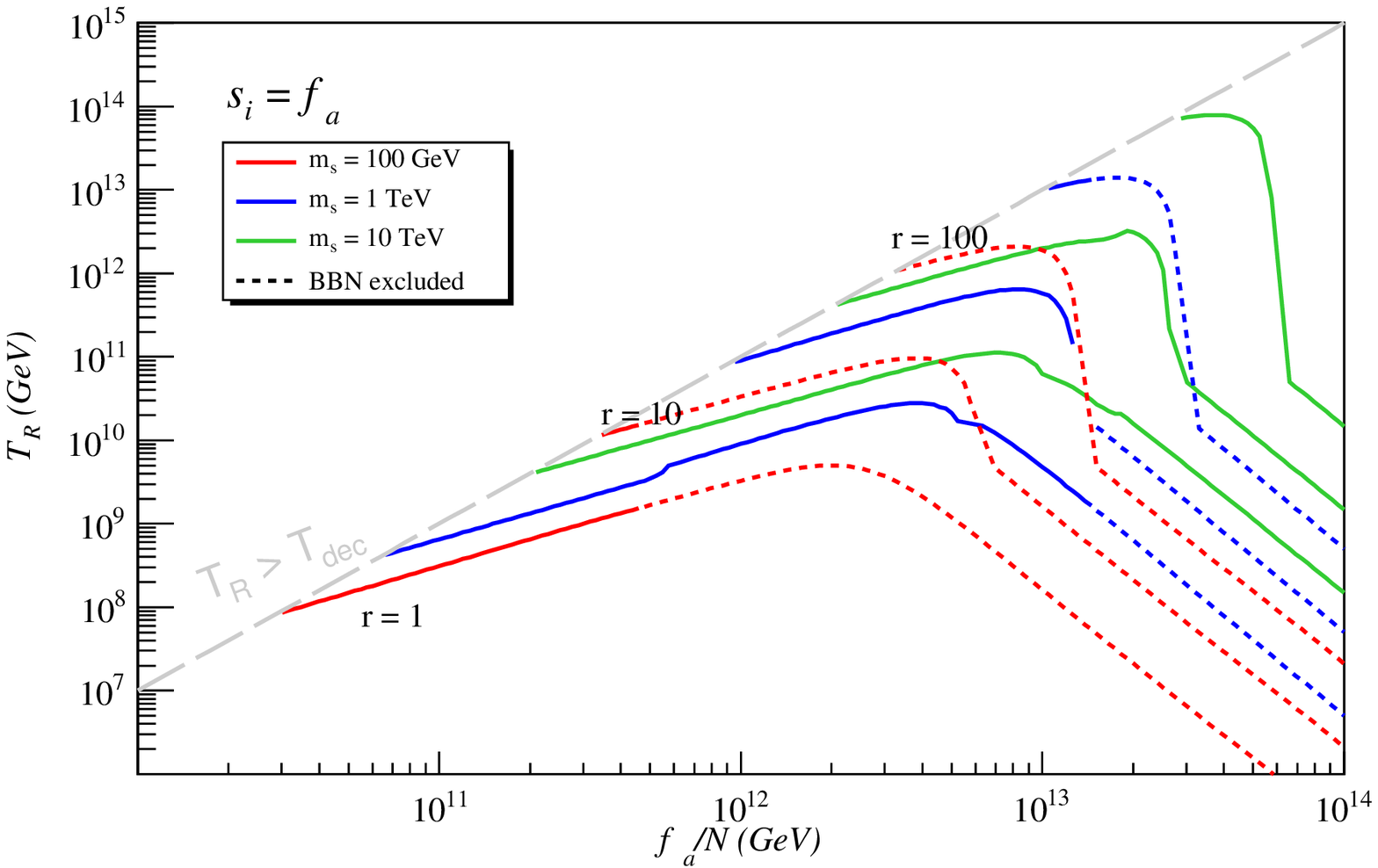}
\includegraphics[width=11cm]{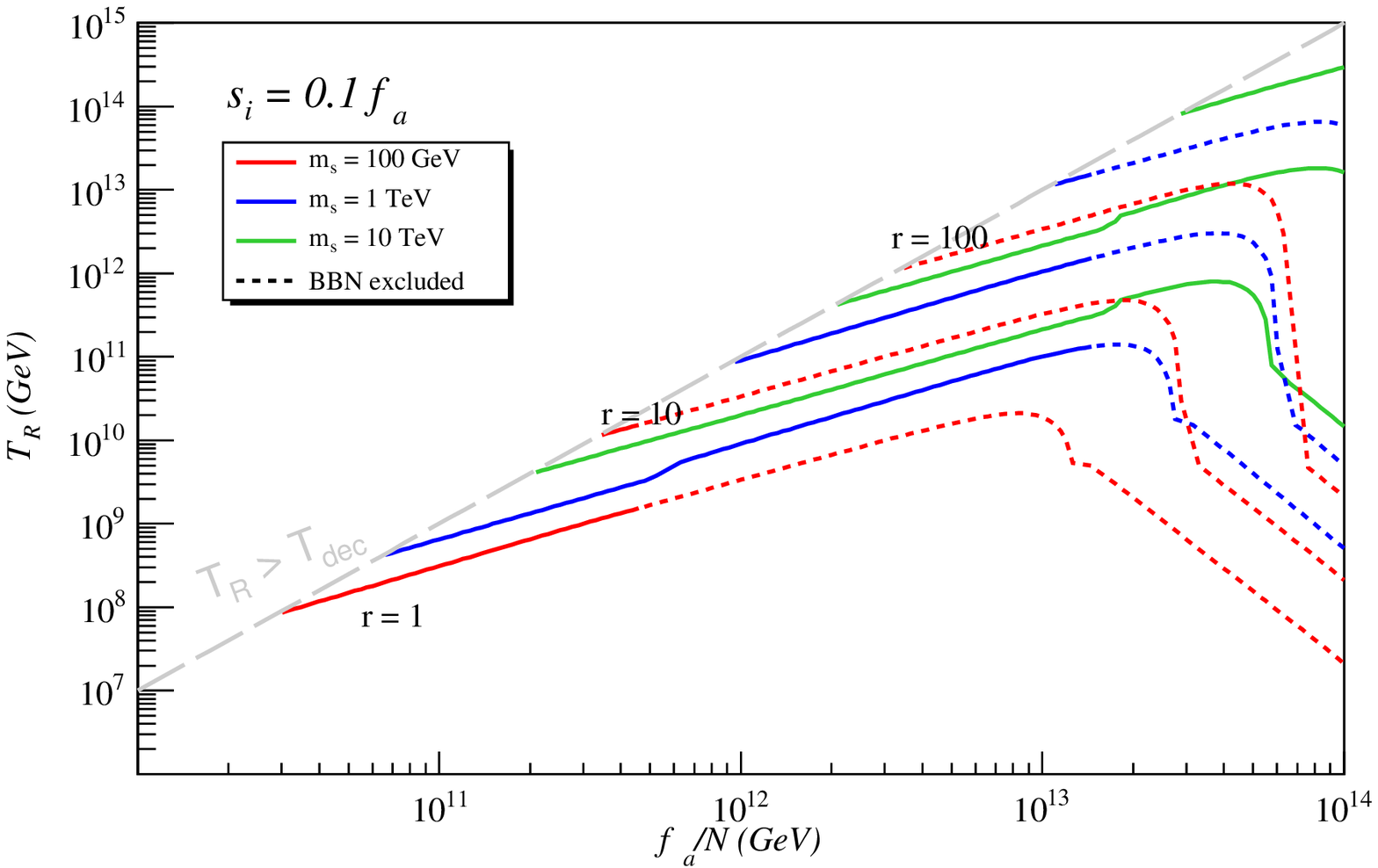}
\caption{Ratio of entropy $r$ before and after saxion decay 
in the $f_a/N\ vs.\ T_R$ plane for $m_s=0.1,\ 1,\ 10$~TeV and for 
{\it a})~$s_i/(f_a/N)=1$ and {\it b})~$s_i/(f_a/N)=0.1$. The dashed
lines correspond to $T_{decay}<10$ MeV, when the entropy from saxion 
decay is injected after the beginning of BBN; these regions are 
likely excluded.
}\label{fig:R}}
%%%%%%%%%%%%%%%%%%%%%%%%%%%%%%%%%%%%%%%%%%%%%%%%%%%%%%%%%%%%%%%%%%%

In Fig.~\ref{fig:R}{\it b}), we plot again the entropy ratio contours, but this time
taking $s_i/(f_a/N)=0.1$. In this case, saxion production from coherent oscillations
is suppressed by the smaller initial saxion field strength value. This 
expands the range of large $T_R$ at high $f_a/N$ where entropy injection is negligible.
In cases such as these, the results of the previous section 
(and also Ref. \cite{Baer:2010kw}) remain viable, and 
entropy injection from saxion decay would be a negligible effect. 
From here on, we will assume $s_i/(f_a/N)=1$.

As mentioned before, the entropy injection from late decaying saxions will
dilute the number density of any particle decoupled from the thermal
plasma at $T = T_s$. Therefore, depending on $T_s$, the saxion production 
and decay may dilute thermally produced axinos, gravitinos,
the quasi-stable $\tz_1$s, sometimes the axions and of course the
matter-antimatter asymmetry itself. In this latter case, the
baryon-to-photon ratio $\eta_B\equiv n_B/n_\gamma$ is diluted to a 
value $\eta_B/r$.

To accommodate the latter possibility, we note that the lepton asymmetry and
consequently the baryon assymetry is proportional to the mass of the lightest
right-handed neutrino, $M_1$ (assuming $M_1\ll M_{2,3}$)\cite{epsilon1}. Therefore,
to compensate the saxion dilution, heavier neutrinos are necessary. Since
thermal leptogenesis requires $T_R > M_1$, in the case of saxion entropy injection,
we need $T_R > r\times(2\times10^{9})$ GeV.

%To accommodate the latter possibility, we note that the lepton asymmetry
%parameter $\epsilon_1$ is calculated to be, under the assumption of
%one lightest of the three heavy neutrinos 
%$M_1\ll M_{2,3}$\cite{epsilon1}
%
%\be
%\epsilon_1\simeq -\frac{3}{16\pi}\frac{M_1}{(hh^\dagger)_{11}v_F^2}
%Im\left(h^*m_{\nu}h^\dagger\right)_{11} ,
%\ee
%
%where $h$ is the lepton sector Yukawa coupling matrix, $v_F$ is the 
%electroweak vev and $M_1$ is the lightest of the singlet right-hand neutrinos.
%The lepton asymmetry yield is then given by
%\be
%\frac{n_L-n_{\bar L}}{s}=-\kappa \frac{\epsilon_1}{g_*}
%\ee
%with $s$ the entropy density, $\kappa$ is a washout factor and $g_*$ is 
%degrees of freedom as usual. Since the ultimate baryon asymmetry 
%is proportional to the lepton asymmetry, 
%if we dilute the baryon asymmetry by a factor $r$, 
%we will need to compensate by increasing the lepton asymmetry by a factor $r$.
%This can be done by requiring $M_1$ increased by a factor $r$, and hence
%the value of re-heat temperature wil need to be increased from a minimum
%$T_R\sim 2\times 10^9$ GeV to $T_R\agt 2r\times 10^9$ GeV.
%To include the effect of the needed increase in re-heat temperature
%under dilution from saxions, we will in this section calculate values of 
%$T_R/r$ and require 
%\be 
%T_R/r\agt 2\times 10^9\ {\rm GeV}.
%\ee

To include the above effects into our new analysis, 
we adopt the following procedure:
\bi
\item Calculate the thermal plus coherent oscillation yield of saxions $Y_s$
in the early universe. 
\item Calculate the saxion decay temperature $T_s$.
\item Calculate $T_e$ and determine if saxions can dominate the universe ($T_e>T_s$).
\item Calculate the final/initial entropy ratio $r$.
\item If $r<1$, then the entropy injection is negligible and we require the saxion lifetime and relic density to satisfy the BBN bounds
for late decaying particles in a radiation dominated universe,
\item If $T_s<10$ MeV and $r>1$, 
then the point is excluded due to entropy injection during or after BBN.
\item If $T_s>10$ MeV and $r>1$,
\bi
\item dilute thermally produced axinos by factor $r$.
\item dilute thermally produced gravitinos by $r$.
\item If $T_s< T_{QCD} = 1$~GeV, dilute mis-alignment produced axions by $r$.
\item If  $T_s<T_{fo} = m_{\tz_1}/25$, dilute quasi-stable neutralinos by $r$.
This condition can dilute axinos produced by neutralino decay, but also impacts the
quasi-stable neutralino BBN bounds from Fig.~\ref{fig:bbn}.
\ei
\ei

Our first results are shown in Fig.~\ref{fig:Oh2R} for the same PQMSSM parameters used
in Fig.~\ref{fig:Oh2_SUG2}, but including saxion entropy injection with $m_s = 1$~TeV.
In frame {\it a}), we plot the relic abundance of thermally produced axinos (red), axions (blue),
 gravitino produced axinos (lavender) and neutralino produced axinos (magenta).
The value of $T_R$ is always adjusted to maintain
$\Omega_{a\ta}=0.1123$, and so $T_R/r$ 
is shown in frame {\it b}) for $m_{\ta}=0.1$ and 1 MeV. 

For low values of $f_a/N$, the relic abundance curves track
the values shown in Fig.~\ref{fig:Oh2_SUG2}. In this case, $T_R$ is much lower than
the leptogenesis value $2\times 10^9$~GeV, and the thermal yield of saxions  is 
too low for significant entropy production. 
As $f_a/N$ increases, the thermal axino production drops, and the value of $T_R$ 
must compensate by increasing the thermal yield of axinos so that
$\Omega_{a\ta}h^2=0.1123$ is maintained.
At around $f_a/N \sim 10^{13}$~GeV, the value of $T_s$ drops below $T_e$, and
significant entropy production from saxion decay occurs. 
The entropy injection dilutes the thermal axino and also axion production, so that
a sharp increase in $T_R$ is needed to offset the dilution effect: the
dark matter abundance remains dominated by thermal axino production.
However, the axion abundance is independent of $T_R$, and so its dilution due to 
saxion decay is plain to see in frame {\it a}). 
Since the entropy injection from
saxion decay also dilutes the matter-antimatter asymmetry, we also show the
trajectory of $T_R/r$ once entropy injection is started. 
%
%%%%%%%%%%%%%%%%%%%%%%%%%%%%%%%%%%%%%%%%%%%%%%%%%%%%%%%%%%%%%%%%%%%
\FIGURE[t]{
\includegraphics[width=10cm]{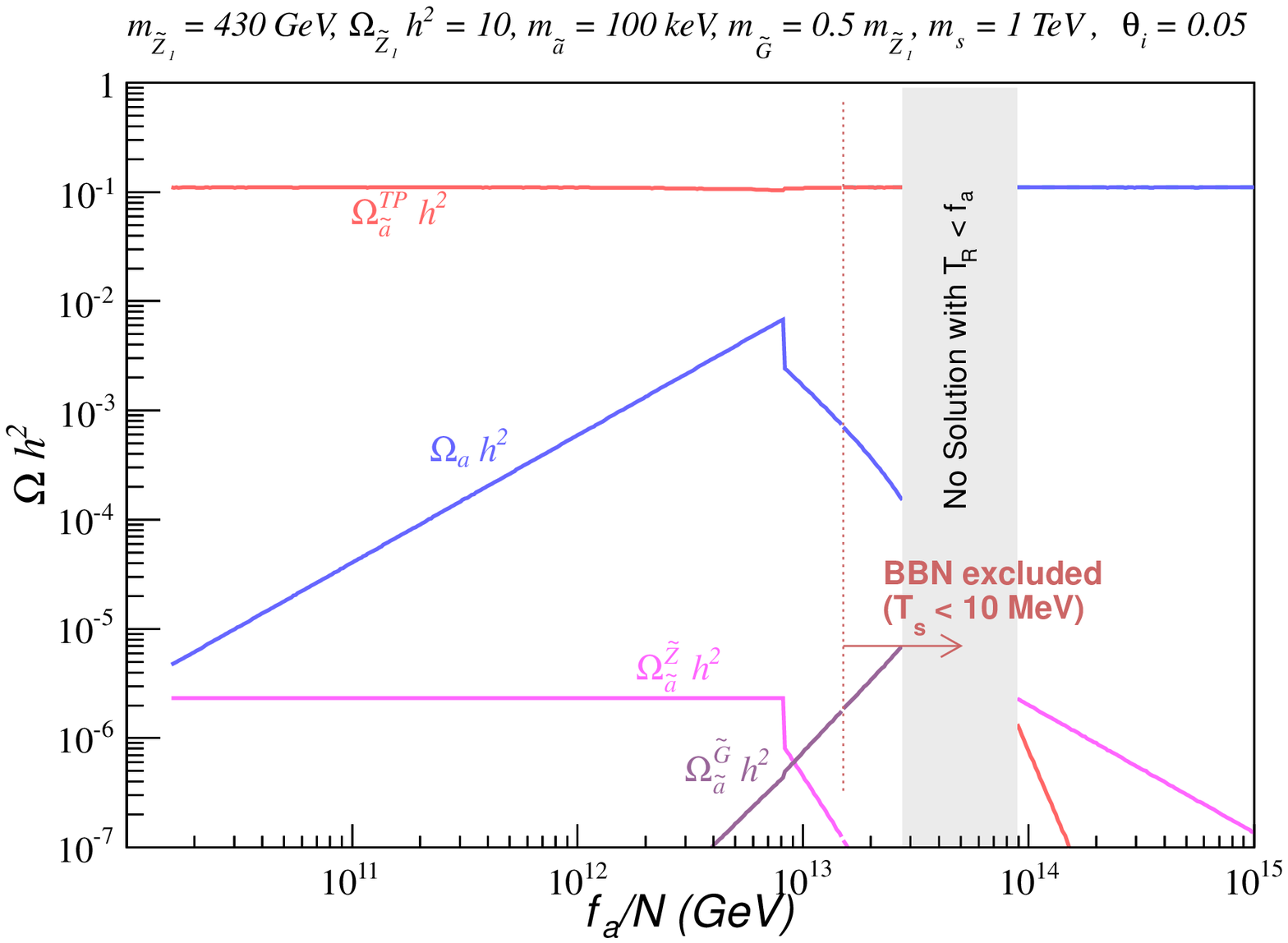}
\includegraphics[width=10cm]{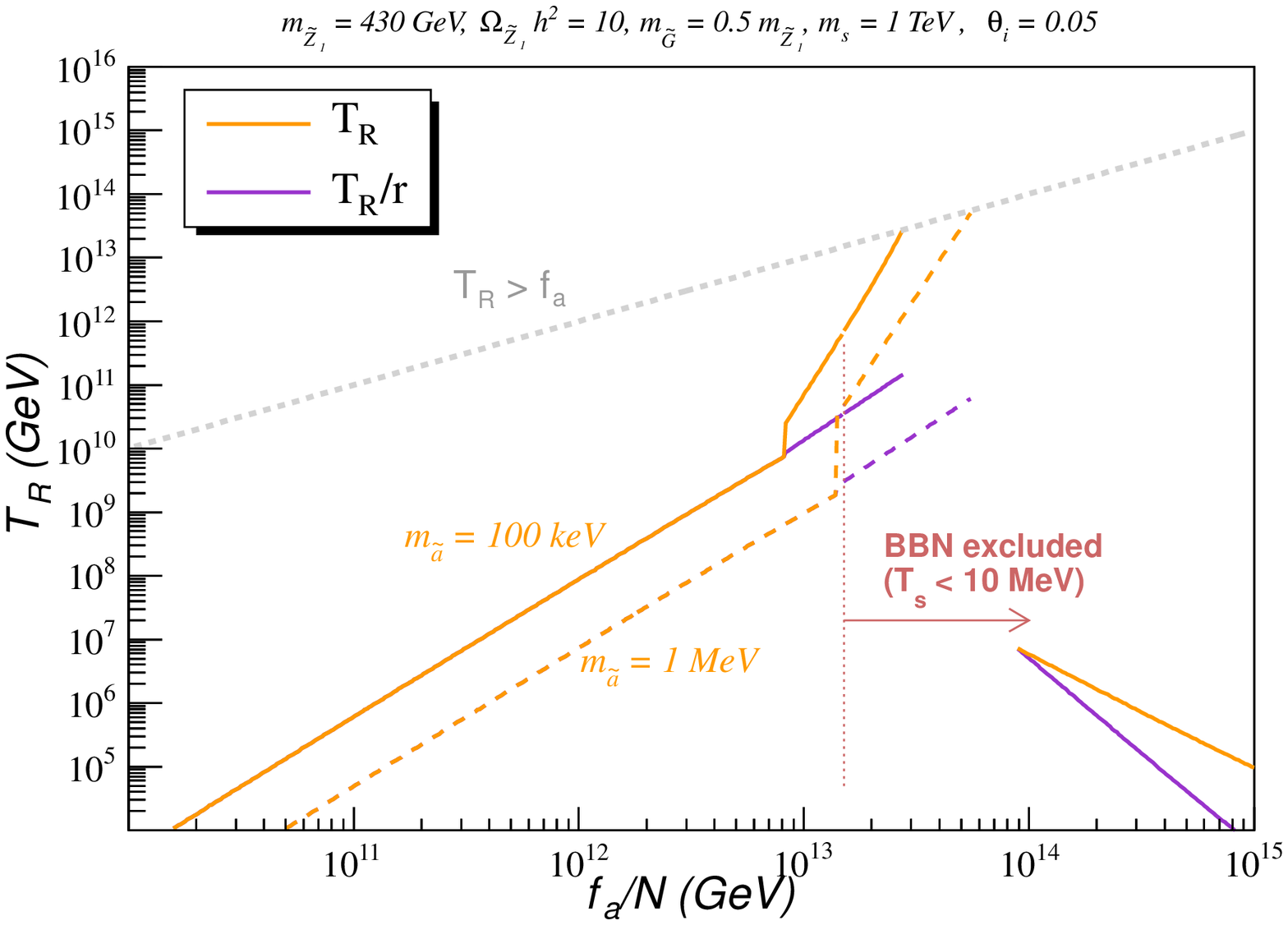}
\caption{Upper frame a): Contribution of axions and TP and NTP  
axinos to the DM density as a function of the PQ breaking scale $f_a/N$, 
for $\Omega_{\tz_1} h^2 = 10$ and $m_{\tz_1} = 430$~GeV, with
 $m_{\ta}=100$~keV and $\theta_i=0.05$; $T_R$ is adjusted such that 
$\Omega_{a\ta}h^2=0.1123$. We assume $m_s=m_0 = 1$~TeV and $m_{\tG}=m_{\tz_1}/2$.
Lower frame b): the value of $T_R$ (and $T_R/r$) 
that is needed to achieve $\Omega_{a\ta} h^2 = 0.1123$ 
for $m_{\ta} = 0.1$ and 1 MeV.
}\label{fig:Oh2R}}
%%%%%%%%%%%%%%%%%%%%%%%%%%%%%%%%%%%%%%%%%%%%%%%%%%%%%%%%%%%%%%%%%%%

The value of $T_R$ needed to maintain $\Omega_{a\ta}h^2=0.1123$ increases sharply 
until the regime $T_R>f_a/N$ is reached. For such high values of $T_R$, 
the PQ symmetry is restored during re-heat, and re-broken during subsequent cooling.
The universe should break into domains of different $\theta_i$ and $s_i$ values (see {\it e.g.} 
M. Turner in Ref. \cite{vacmis}), and a modified treatment of dark matter will be needed. Therefore
we neglect such solutions and impose the condition $T_R < f_a/N$ to our solutions.

As $f_a/N$ increases even further, we move into the range where $T_s<10$ MeV,
and saxion decay might disrupt BBN. 
In this excluded region, two solutions to the restriction
$\Omega_{a\ta}h^2=0.1123$ appear. The first has dark matter dominated by 
thermal axinos and ultra-high $T_R\gg f_a/N$ wherein
the axinos and axions are 
severely diluted by saxion entropy production; these solutions are not exhibited on the plot.
The second solution allows for much lower $T_R$ values in which case dark matter
is dominated by axion production, albeit with some dilution due to coherent
oscillation production of saxions.
These high $f_a/N$ solutions, however intriguing, are all excluded because such high values
of $f_a/N$ suppress the saxion and $\tz_1$ lifetimes, so their decays will affect BBN.

The upshot of Fig.~\ref{fig:Oh2R} is that, for $f_a/N$ slightly below $10^{13}$~GeV, 
the value of $T_R$ has increased to over $10^{11}$~GeV while maintaining
$\Omega_{a\ta}h^2=0.1123$. Although the saxion entropy injection leads to higher
values of $T_R$ (when compared to Fig.\ref{fig:Oh2_SUG2}), the allowed range for the
relevant temperature for leptogenesis ($T_R/r$) is actually reduced, due to the BBN bounds on the decaying saxion.
For the case of $m_{\ta}=0.1$ MeV, 
the value of $T_R/r$ reaches to $\sim 10^{10}$ GeV for $f_a/N\sim 10^{13}$ GeV, 
thus potentially reconciling thermal leptogenesis with the gravitino problem 
in the AY scenario.
However, the solution with $m_{\ta}=1$ MeV never reaches as high as $T_R/r\sim 2\times 10^9$ GeV
before entering the BBN-excluded region, and so in this case does {\it not} 
lead to a reconciliation of themal leptogenesis with the gravitino problem.
Comparing Figs. \ref{fig:Oh2_SUG2} and \ref{fig:Oh2R} we can see that, at least for this case, 
the range of $f_a$ values which accomodates thermal leptogenesis is actually {\it reduced} when the
saxion entropy injection effect is included.

\subsection{Scan over PQMSSM parameters including dilution due to entropy injection from saxion decay}
\label{sec:rscan}

While Fig.~\ref{fig:Oh2R} holds for particular values 
of the PQMSSM parameters, save for $f_a/N$, we will now scan over the remaining PQ 
parameters $m_{\ta}$ and $\theta_i$, as well as $f_a/N$, as in 
Section~\ref{sec:scan}.
%Fig.~\ref{fig:sug2fa}.
This time, we will adopt $m_s=m_0=1$~TeV and $s_i/(f_a/N)=1$, and allow for
saxion-induced entropy dilution of mixed axion/axino DM according to the procedure
described in the last section.  
The results for $\Omega_{\tz_1} h^2 = 0.04$ and $m_{\tz_1} = 430$~GeV
 are shown in Fig.~\ref{fig:sug2faR}, where we
plot the value of $T_R/r$ needed to maintain $\Omega_{a\ta}h^2=0.1123$ versus
PQ breaking scale $f_a/N$. The line where $r=S_f/S_i =1$ is shown in magenta.
The red points violate BBN bounds due to late decaying $\tz_1$, while the green
points violate BBN bounds due to late-time saxion decays ($T_s < 10$ MeV).
The light blue points have $>20\%$ WDM or $>1\%$ HDM, while
the dark blue points satisfy all constraints. 
The WDM/CDM bound following Boyarsky\cite{warm} is again indicated as a dashed blue line. 
By including dilution of DM from saxion production and decay,
the allowed points can reach to $T_R$ as high as $\sim 10^{13}$~GeV for 
$f_a/N\sim 1.5\times 10^{13}$~GeV although the value of $T_R/r$ reaches only
as high as $\sim 10^{11}$ GeV. These points with $T_R/r\agt 2\times 10^9$ GeV 
evidently reconcile thermal leptogenesis with the gravitino problem
even in the presence of entropy injection from saxion decay.

%%%%%%%%%%%%%%%%%%%%%%%%%%%%%%%%%%%%%%%%%%%%%%%%%%%%%%%%%%%%%%%%%%%
\FIGURE[t]{
\includegraphics[width=10cm]{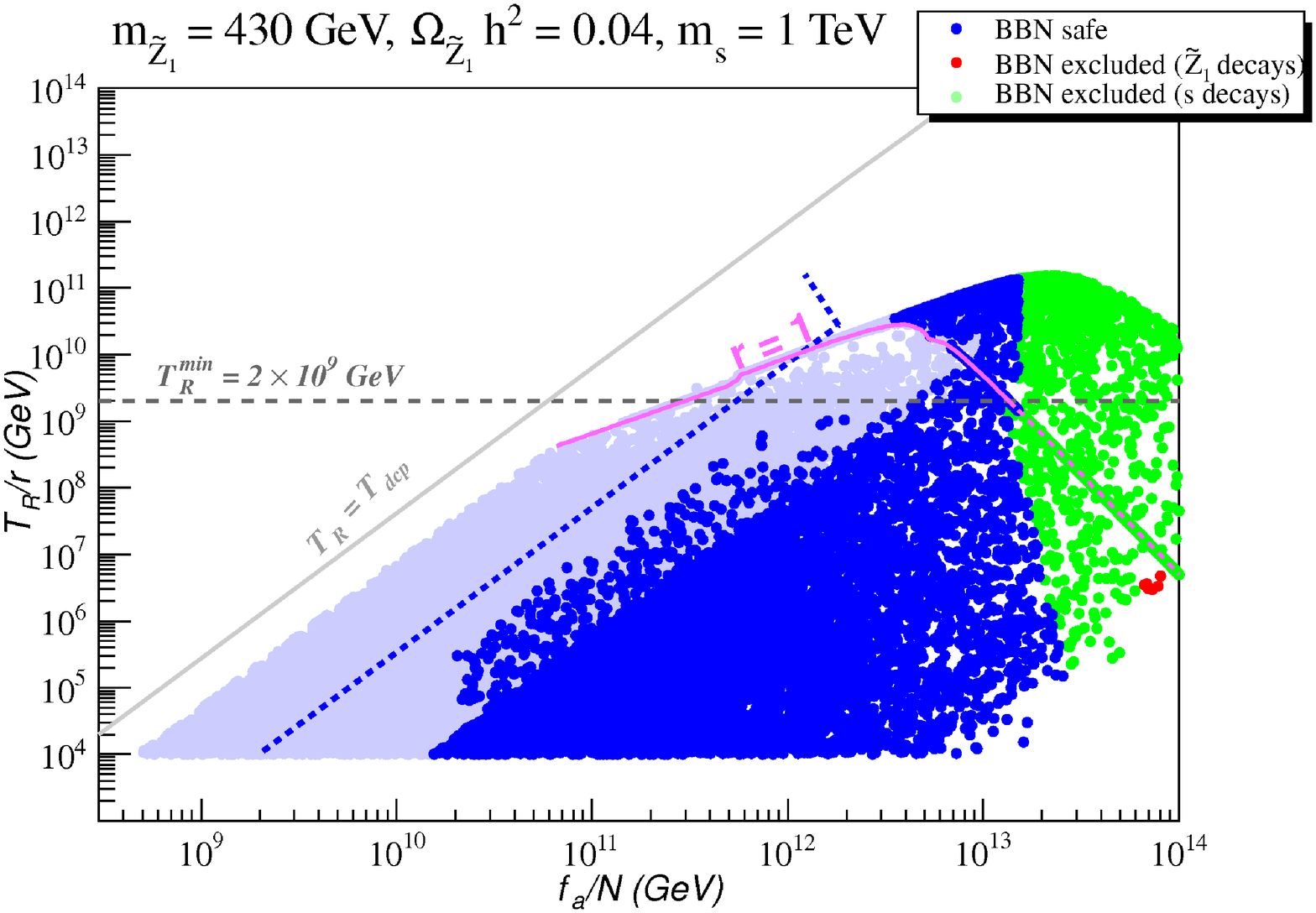}
\caption{Allowed and disallowed points in the $f_a\ vs.\ T_R/r$ plane
for $\Omega_{\tz_1} h^2 = 0.04$ and $m_{\tz_1} = 430$~GeV, with $m_s= 1$~TeV.
}\label{fig:sug2faR}}
%%%%%%%%%%%%%%%%%%%%%%%%%%%%%%%%%%%%%%%%%%%%%%%%%%%%%%%%%%%%%%%%%%%
%

In Fig.~\ref{fig:sug2thR}, we plot the axion mis-alignment angle $\theta_i$. 
Unlike the previous results in Fig.~\ref{fig:sug2th} with no entropy injection, 
the allowed values of $\theta_i$ with $T_R/r >2\times 10^9$~GeV
span a range from $0$ to $\sim 1.5$ radians: for higher values of
$T_R$, larger values of $\theta_i$ can be tolerated since the
relic abundance of axions is now diluted by saxion decay. Therefore,
in this case, the axion mis-alignment angle is not required
to take unnaturally small values, as opposed to the case without the saxion
dilution, shown in Fig.\ref{fig:sug2fa}. 
%
%%%%%%%%%%%%%%%%%%%%%%%%%%%%%%%%%%%%%%%%%%%%%%%%%%%%%%%%%%%%%%%%%%%
\FIGURE[t]{
\includegraphics[width=10cm]{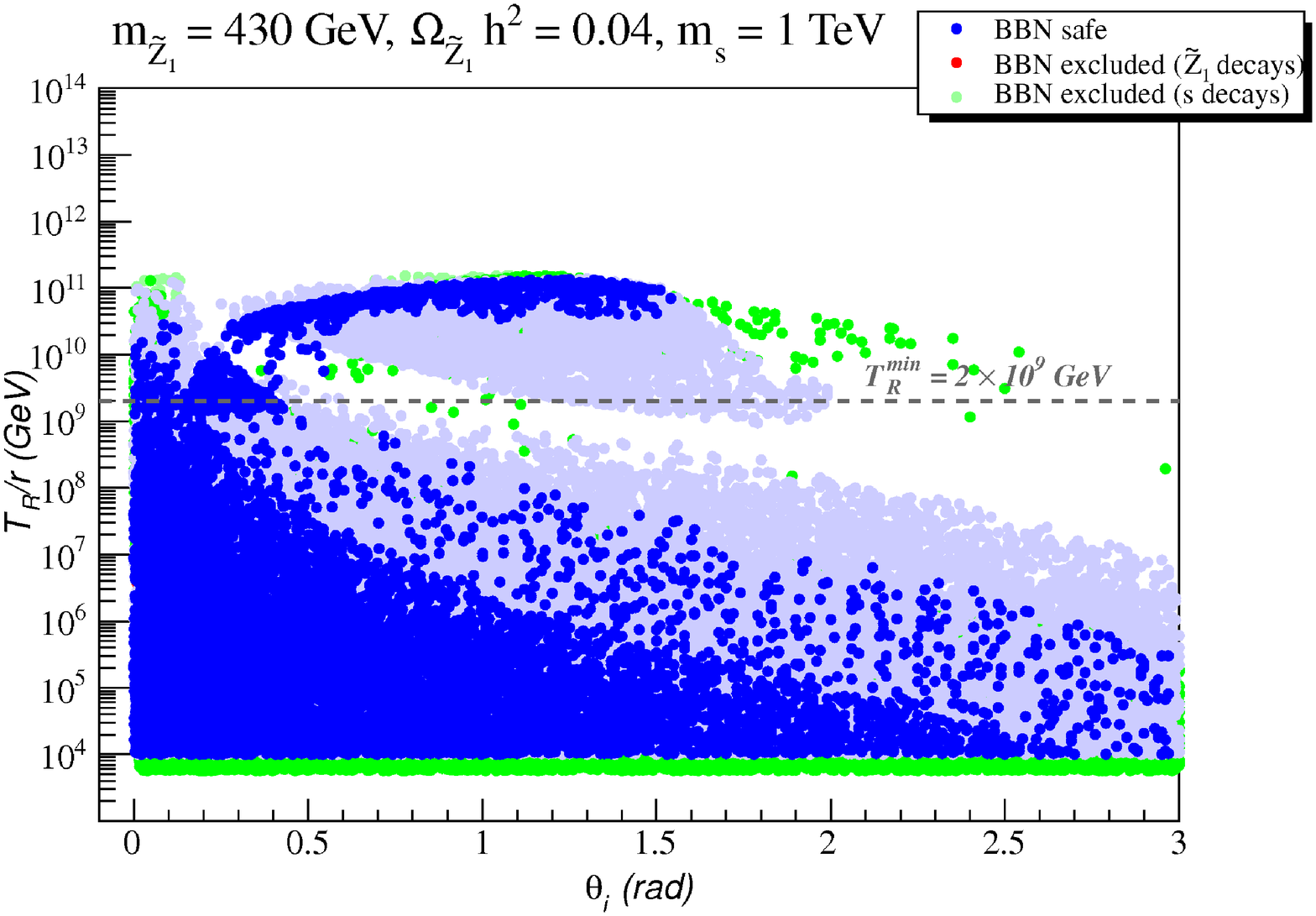}
\caption{Allowed and disallowed points in the $\theta_i\ vs.\ T_R/r$ plane
for $\Omega_{\tz_1} h^2 = 0.04$ and $m_{\tz_1} = 430$~GeV, with $m_s= 1$~TeV.
}\label{fig:sug2thR}}
%%%%%%%%%%%%%%%%%%%%%%%%%%%%%%%%%%%%%%%%%%%%%%%%%%%%%%%%%%%%%%%%%%%
%

To see whether axinos or axions dominate the DM density
including entropy from saxions, in Fig.~\ref{fig:sug2axR} we plot the same points, but this time versus
axion relic density $\Omega_ah^2$. We see that the bulk of points with
$T_R/r >2\times 10^9$~GeV that are BBN-allowed indeed have {\it mainly axion} CDM. 
Note that the point shown in Fig.~\ref{fig:Oh2R}, which has $\theta_i = 0.05$
and mainly axino DM (at $T_R > 2\times 10^9$~GeV),
corresponds to the few points of Fig.~\ref{fig:sug2axR}
at low $\Omega_ah^2$ and is not the most common scenario, since it requires
quite small values of the mis-alignment angle.
Given that $f_a/N\sim 3-15\times 10^{12}$~GeV, we expect the axion mass 
$m_a\sim 0.4-2$ $\mu$eV, somewhat below the range where ADMX is searching\cite{admx}.
%%%%%%%%%%%%%%%%%%%%%%%%%%%%%%%%%%%%%%%%%%%%%%%%%%%%%%%%%%%%%%%%%%%
\FIGURE[t]{
\includegraphics[width=10cm]{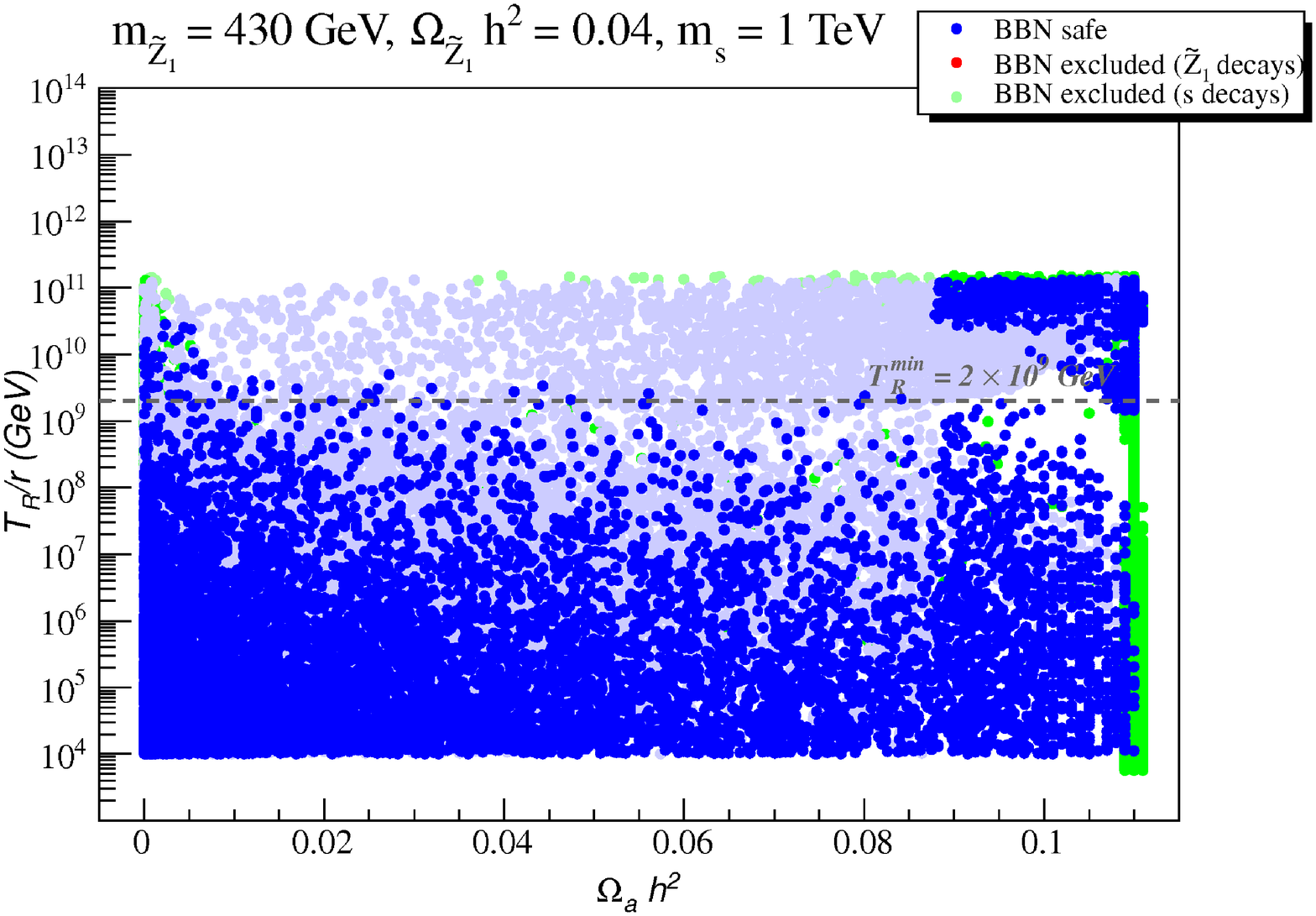}
\caption{Allowed and disallowed points in the $\Omega_a h^2\ vs.\ T_R/r$ plane
for $\Omega_{\tz_1} h^2 = 0.04$ and $m_{\tz_1} = 430$~GeV, with $m_s= 1$~TeV.
}\label{fig:sug2axR}}
%%%%%%%%%%%%%%%%%%%%%%%%%%%%%%%%%%%%%%%%%%%%%%%%%%%%%%%%%%%%%%%%%%%

\subsection{More general scan over MSSM parameters}
\label{sec:scanR}

Next, we generalize our results for a general PQMSSM model, where we now
allow $\Omega_{\tz_1}$ and $m_{\tz_1}$ to be free parameters included in our
scan, as in Fig.~\ref{fig:OmMZ1}. For simplicity we keep the saxion mass
fixed at $m_s = 1 $~TeV. We keep only points with $T_R/r>2\times 10^9$ GeV, which
potentially reconcile thermal leptogenesis with the gravitino problem.
The result is shown in Fig.~\ref{fig:OmMZ1R}, where the red  points are
excluded due to the BBN constraints on $\tz_1$ decays; no green points 
due to constraints from BBN on saxion decay are visible. 
By comparing Figs. \ref{fig:OmMZ1} and \ref{fig:OmMZ1R}, 
we see that due to the saxion dilution of the neutralino relic density,
the BBN bounds on $\Omega_{\tz_1}$ are less severe 
and a larger portion of the MSSM parameter space can be consistent
with thermal leptogenesis. 
%On the other hand, the PQ parameter space
%becomes more constrained due to the BBN bounds on saxion decays,
%which are shown as green points in Fig.~\ref{fig:OmMZ1R}. 
%As a consequence, the density of CDM/BBN consistent solutions with 
%$T_R/r > 2 \times 10^9$~GeV (dark blue dots)
%is clearly smaller than in Fig.~\ref{fig:OmMZ1}. 
%
%%%%%%%%%%%%%%%%%%%%%%%%%%%%%%%%%%%%%%%%%%%%%%%%%%%%%%%%%%%%%%%%%%%
\FIGURE[t]{
\includegraphics[width=10cm]{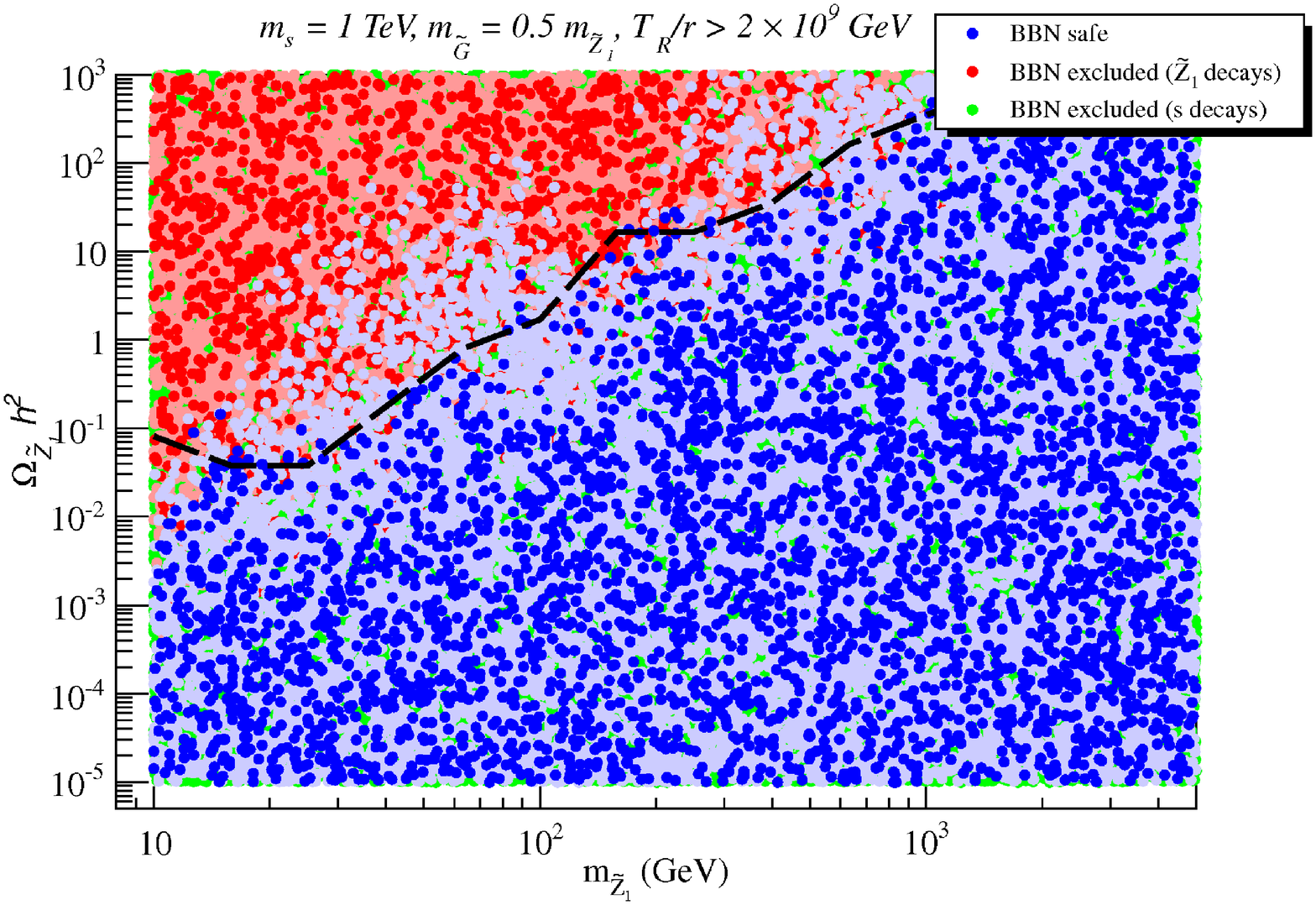}
\caption{Allowed and disallowed points in the $\Omega_{\tz_1} h^2\ vs. m_{\tz_1}$ plane
for a general scan over SUSY models with a bino $\tz_1$ and $m_s = 1$~TeV.
For all points, we require $T_R/r >2\times 10^9$~GeV.
In this plot, we include the effect of entropy production from saxion
decay. Dark blue points are consistent with BBN and have mainly CDM with at 
most 20\%~WDM and/or 1\%~HDM admixture. 
%The region below the dashed line defines for each value of ${\tz_1}$, the $\Omega_{\tz_1} h^2$ values below which 99\% of the CDM/BBN consistent solutions lie. 
The region below the dashed line represents the MSSM parameter space
where 99\% of the DM/BBN consistent solutions lie when applying weaker 
WDM/HDM requirements as discussed in the text.
}\label{fig:OmMZ1R}}
%%%%%%%%%%%%%%%%%%%%%%%%%%%%%%%%%%%%%%%%%%%%%%%%%%%%%%%%%%%%%%%%%%%
%

\subsection{Thermal leptogenesis-allowed regions of the mSUGRA plane}
\label{ssec:msugra}

As a last point of this study, let us apply our general results to 
the showcase mSUGRA model in the $m_0\ vs.\ m_{1/2}$ plane. 
In order to make our results independent of a particular choice 
of PQ parameters, we consider the bounds on $\Omega_{\tz_1} h^2$ obtained
from the general PQMSSM scan in Sections~\ref{sec:scan} and \ref{sec:scanR}
(for the case with saxion entropy injection). These bounds are represented
by the dashed lines in Figs.~\ref{fig:OmMZ1} and \ref{fig:OmMZ1R}.
We may then translate this into a contour in the $m_0\ vs.\ m_{1/2}$
plane of mSUGRA for $A_0=0$, $\mu >0$ and constant $\tan\beta$, 
as shown for the cases of $\tan\beta=10$, $50$ and $55$ in Fig.~\ref{fig:msugra}. 
The gray regions are excluded because they violate the LEP2 limits on Higgs 
or sparticle masses\footnote{The LEP2 limit on a SM-like Higgs scalar $h$ 
is $m_h>114.4$~GeV. Here, we use $m_h>111$~GeV allowing for an approximate 
3~GeV error on the theory calculation of $m_h$. For the SUSY mass limits we 
use those implemented in SuperIso~\cite{superiso}.}
or have a stau as next-to-next-to-lightest SUSY particle (NNLSP); 
this latter case requires special
treatment as for example in Ref.~\cite{freitas}.

In frame {\it a)}, we show the mSUGRA $m_0\ vs.\ m_{1/2}$ plane for $\tan\beta=10$.  
The strips of dark blue and purple points show the regions that allow for 
$T_R>T_R^{min}=2\times 10^9$ GeV, 
while maintaining $\Omega_{a\ta}h^2=0.1123$ and respecting bounds from BBN. 
The subset of purple points at low $m_{1/2}$ satisfies in addition  
the following constraints on low energy (LE) observables:%~\cite{pdb}:
\begin{enumerate}
\item $\Delta a_\mu^{SUSY} = (7.90 - 37.39)\times 10^{-10}$ ,
\item $BR(b\to s\gamma ) = (2.79 - 4.3)\times 10^{-4}$,
\item $BR(B_s\to \mu^+\mu^- ) < 4.7\times 10^{-8}$,
\item $0.55<BR(B_u\to\tau^+\nu_\tau )^{\rm MSSM}/BR(B_u\to\tau^+\nu_\tau )^{\rm SM}<2.71$
\end{enumerate}
where $1.-3.$ were calculated using Isajet/Isatools and 
$4.$ was calculated using SuperIso.  

We see that the AY consistent regions, although broader, are very similiar to 
the classic mSUGRA regions with neutralino dark matter: the stau co-annihilation 
region at low $m_0$ and the light Higgs resonance region where $\tz_1\tz_1\to h$ 
at $m_{1/2}\sim150$~GeV. The reason is that a rather low abundance of thermal 
neutralinos is required in the AY scenario to satisfy BBN constraints on late
decaying $\tz_1$s. 
For comparison, the classic mSUGRA strips where the neutralino relic 
density $\Omega_{\tz_1}h^2=0.1123\pm0.0105$ are shown as yellow/orange points. 
 
Invoking next the $\Omega_{\tz_1}\ vs.\ m_{\tz_1}$ contour
of Fig.~\ref{fig:OmMZ1R}, which includes the effect of entropy
generation from a $m_s=1$ TeV saxion, the AY-consistent regions
broaden out considerably. The region with $T_R/r>T_R^{min}$ is
denoted here by light blue points, and expands to fill the lower $m_0$
portion of the $m_0\ vs.\ m_{1/2}$ plane along with a band around 
$m_{1/2}\sim 400$ where turn-on of the $\tz_1\tz_1\to t\bar{t}$ 
annihilation chanel reduces the neutralino abundance. The portion of the
leptogenesis consistent region including saxion decays and LE constraints
is colored in pink, and requires $m_{1/2}\alt 550$~GeV and $m_0\alt 500$~GeV, 
so as to allow for a significant contribution to $(g-2)_\mu$ by light 
charginos and sneutrinos.
The remaining unshaded (white) region of the mSUGRA plane does not
allow for an AY reconciliation of thermal leptogenesis 
with the gravitino problem, with or without saxion decays, 
mainly because the relic density of neutralinos is so large that
the BBN constraints on late decaying $\tz_1$ are violated.
%
%%%%%%%%%%%%%%%%%%%%%%%%%%%%%%%%%%%%%%%%%%%%%%%%%%%%%%%%%%%%%%%%%%%
\FIGURE[t]{
\includegraphics[width=8cm]{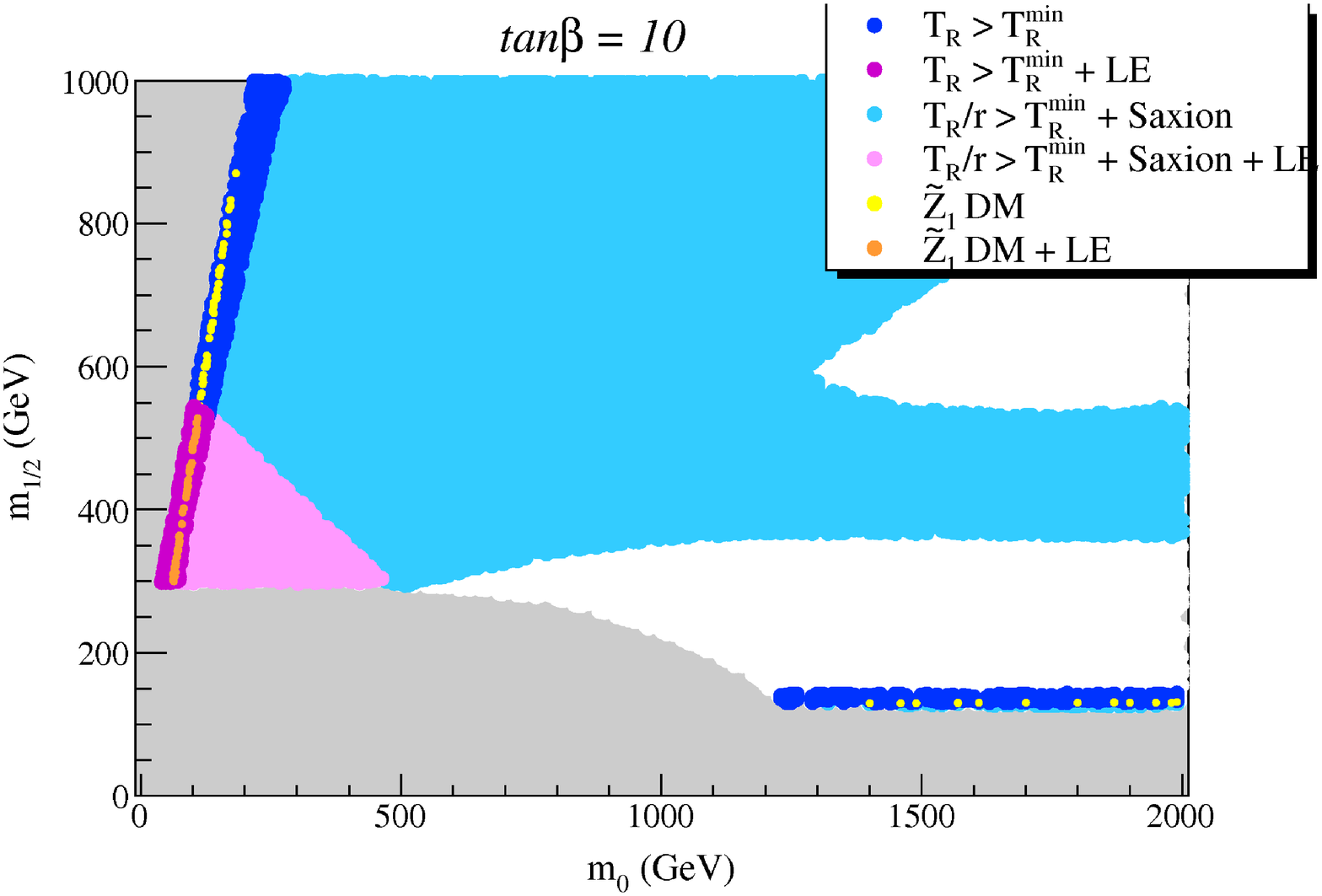}
\includegraphics[width=8cm]{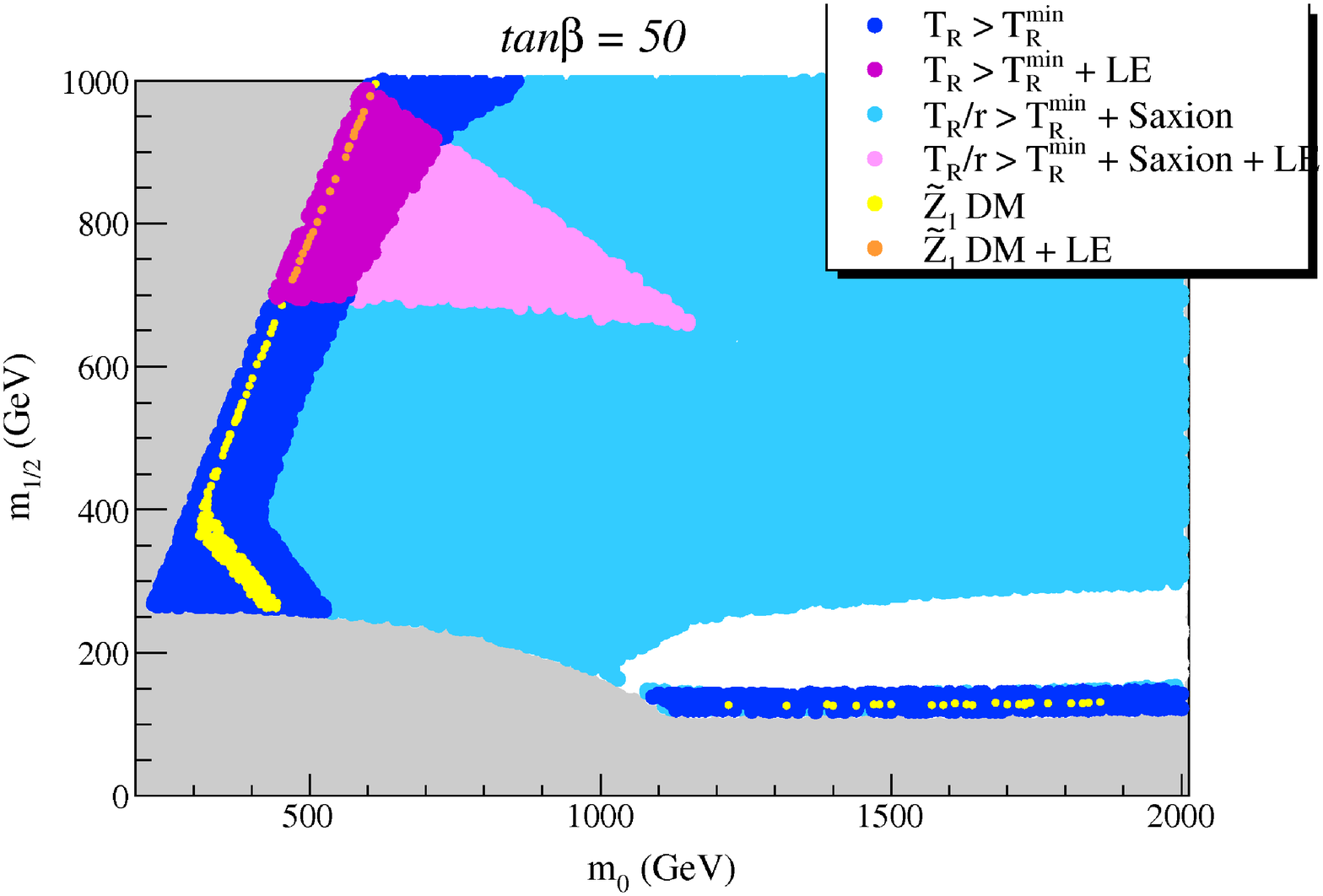}
\includegraphics[width=8cm]{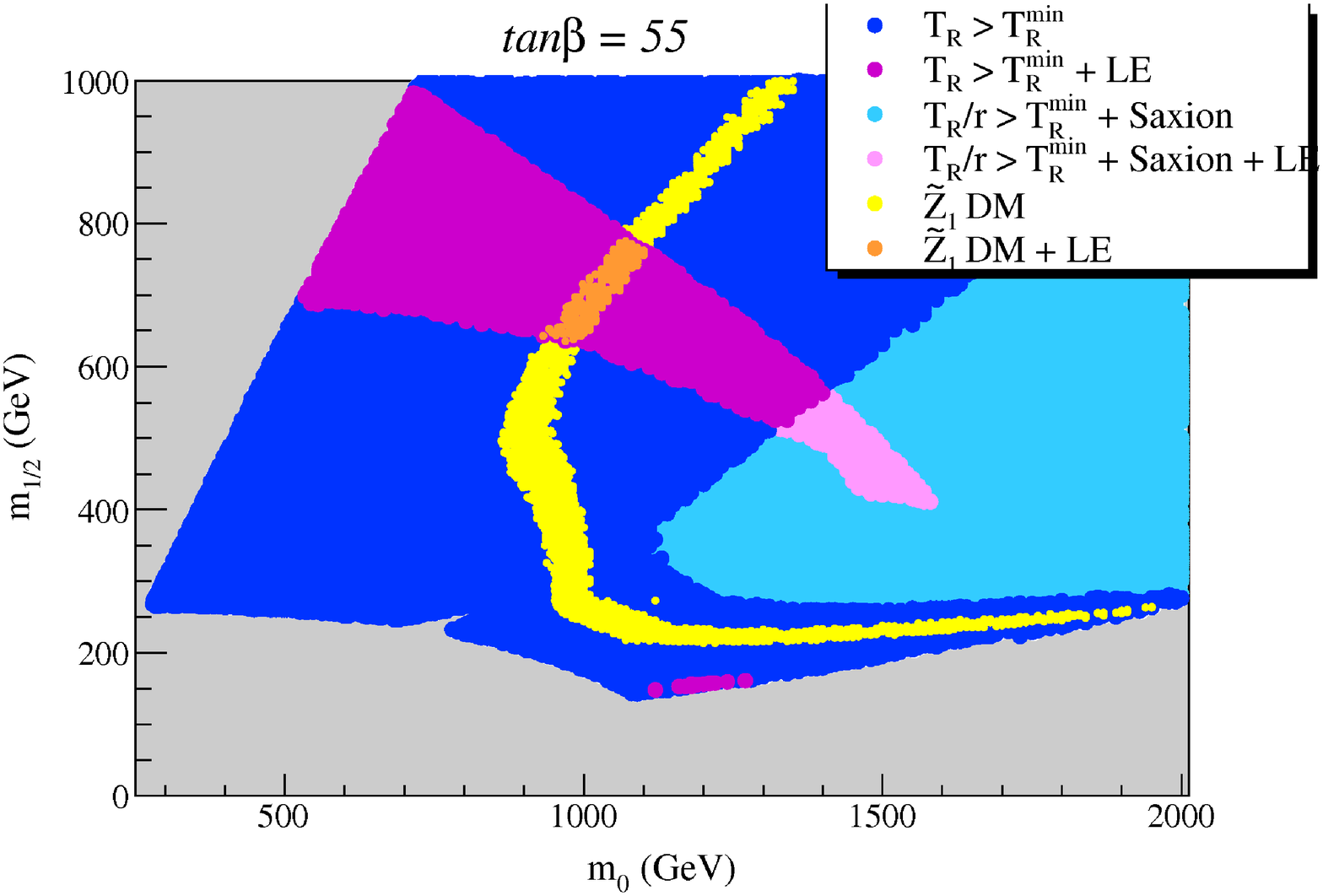}
\caption{Regions in the $m_0\ vs.\ m_{1/2}$ plane of the mSUGRA model with 
$A_0=0$ and $\mu >0$ which satisfy 1.~$T_R>T_R^{min}=2\times 10^9$ GeV (dark blue), 
2.~$T_R>T_R^{min}$ and LE constraints (purple), 
3.~$T_R/r >T_R^{min}$ with saxion entropy injection (light blue) and 
4.~$T_R/r >T_R^{min}$ with saxion entropy injection and LE constraints (pink).
For comparison, the yellow/orange points indicate the classic mSUGRA 
regions with $\Omega_{\tz_1}h^2=0.1123\pm0.0105$.  
We show frames for {\it a})~$\tan\beta =10$, {\it b})~$\tan\beta =50$ and
{\it c})~$\tan\beta =55$.
}\label{fig:msugra}}
%%%%%%%%%%%%%%%%%%%%%%%%%%%%%%%%%%%%%%%%%%%%%%%%%%%%%%%%%%%%%%%%%%%
%

Frame {\it b}) of Fig.~\ref{fig:msugra}, shows the analogous plot for
$\tan\beta =50$. In this case, $b$- and $\tau$-Yukawa couplings increase 
greatly, while the value of $m_A$ drops, enabling efficient annihilation 
of neutralinos via stau coannihilation or $s$-channel $A$ exchange. 
The apparent neutralino abundance $\Omega_{\tz_1} h^2$ is severely reduced, 
and less constrained by BBN.
The area of leptogenesis-consistent regions increases. Furthermore, the SUSY
contributions to $b\to s\gamma$ and $(g-2)_\mu$ increase with increasing $\tan\beta$,
and so the region which is consistent with LE constraints moves to higher $m_{1/2}$
values. If saxion entropy production is added, almost 
the whole plane is allowed by the AY scenario.

Finally, frame {\it c}) shows the case of $\tan\beta =55$, 
where the $A$-resonance dominates the $\tz_1\tz_1$ annihilation amplitudes. 
Here, we see that a huge swath of parameter space is AY-consistent, even without 
the effect of saxion decays. By including entropy from saxion decay, the entire 
$m_0\ vs.\ m_{1/2}$ plane becomes AY-consistent. 
The part which is consistent with LE constraints follows suit, leading to  
a large region of parameter space that is consistent with all constraints.

%========================================================================================
\section{Conclusions}
\label{sec:conclude}
%========================================================================================

In this paper, we reported on investigations of the viability of the Asaka-Yanagida
suggestion that a mass hierarchy with $m({\rm sparticle})>m_{\tG}>m_{\ta}$ can be used
to reconcile thermal leptogenesis, which requires $T_R\agt 2\times 10^9$~GeV, with 
the gravitino problem, which seemingly requires much lower $T_R$ to avoid
BBN constraints and overproduction of neutralino dark matter.
In the AY scenario, the $\tG$ decays inertly to $a\ta$. BBN constraints on
${\rm sparticle}\to \tG +{\rm particle}$ can be avoided because the much faster decays
${\rm sparticle}\to \ta + {\rm particle}$ are now allowed.
We re-examined the AY scenario in Sec. \ref{sec:AY} by including 
1.~updated measurements on the total dark matter abundance $\Omega_{DM}h^2\simeq 0.1123$, 
2.~updated calculations of thermal axino and gravitino production, 
3.~the contribution of relic axions and
4.~BBN constraints on late decaying $\tz_1$s. 
Furthermore, in Sec.~\ref{sec:saxion}, we 
included dilution of dark matter by saxion production and decay. The latter effect can be 
neglected if $m_s$ is in the multi-TeV range and the initial saxion field strength
$s_i$ is somewhat smaller than the PQ breaking scale $f_a/N$.

We found in Sec.~\ref{sec:AY}, neglecting the saxion entropy effect, 
that the AY scenario does work under the conditions that 
{\it (i)}~$f_a/N$ is rather large $\agt 10^{12}$~GeV, implying a somewhat lighter axion 
than is presently searched for by ADMX\cite{admx}, 
{\it (ii)}~the apparent neutralino relic density $\Omega_{\tz_1} h^2$ is not too big:
$\Omega_{\tz_1} h^2\alt 1$, 
{\it (iii)}~the value of $m_{\tz_1}$ is at least in the
several hundred GeV range in order to hasten the $\tz_1$ decay rate, and
{\it (iv)}~the axion mis-alignment angle $\theta_i$ is on the small side $\alt 0.5$ to suppress
overproduction of axions when $f_a/N$ is large.

By including saxion production and decay in Sec. \ref{sec:saxion}, we can dilute the
axino and also axion DM abundance, which in turn allows for somewhat higher values of 
$T_R$ up to $\sim 10^{13}$~GeV to be generated. 
However, since saxion decay also dilutes the baryon density, in this case we must
require instead $T_R/r> 2\times 10^9$ GeV.
The saxion mass $m_s$ needs to be rather large
to avoid BBN constraints on late decaying saxions if $T_R$ is to be high. In this case,
the DM is likely to be {\it mainly axions}, although a few cases with mainly axino DM were 
generated. The axion mis-alignment angle need not be small here since the axion
abundance can be suppressed by entropy injection from saxions. We have also
found that a large portion of the MSSM parameter space ($\Omega_{\tz_1}$ and $m_{\tz_1}$)
can be consistent with high $T_R$ and still avoid the BBN bounds on late decaying
neutralinos, due to the dilution of the neutralino relic density after the entropy
injection from saxion decays.

The observable consequences of our final results are as follows. If the
AY scenario with $m({\rm sparticle})>m_{\tG}>m_{\ta}$ is to reconcile
thermal leptogenesis with the gravitino problem, then we expect several 
broad results to ensue:
\begin{enumerate}
\item discovery of SUSY at the LHC, with a reconstructed
$\Omega_{\tz_1} h^2$ not too large, lest $\tz_1$s are produced at too large a rate
in the early universe, and their late decays disrupt BBN;
\item a SUSY mass spectrum consistent with SUGRA models with a rather light 
(but still weak scale) gravitino, since the gravitino mass must be 
lighter than all observable sparticles;
\item a mainly bino-like $\tz_1$, to quicken decays into $\ta\gamma/Z$, with mass
         $m_{\tz_1}$ in the hundreds of GeV range, which also helps diminish the lifetime;
\item no direct or indirect detection of neutralino (WIMP) dark matter;
\item finally, we expect discovery of an axion to be likely, but in the mass range
$\sim 0.1-2$ $\mu$eV, somewhat below the values presently being explored.
\end{enumerate}

%==============================================================================
\acknowledgments
%==============================================================================

This research was supported in part by the U.S. Department of Energy,
by the Fulbright Program and CAPES (Brazilian Federal Agency for
Post-Graduate Education), and by the French ANR project {\tt ToolsDMColl}, 
BLAN07-2-194882.

% ---- Bibliography ----
%


\begin{thebibliography}{99}
%
\bibitem{nu_review} For some reviews, see {\it e.g.} V. Barger, D.Marfatia and
K. Whisnant, \ijmpe{12}{2003}{569};
L.Camilleri, E. Lisi and J. Wilkerson, \arnps{58}{2008}{343};
B. Kayser, in C.~Amsler {\it et al.}  [Particle Data Group],
%``Review of particle physics,''
\plb{667}{2008}{1}.
%
\bibitem{seesaw}  M. Gell-Mann, P. Ramond and R. Slansky, 
in {\it Supergravity, Proceedings of the Workshop}, Stony Brook, NY 1979 
(North-Holland, Amsterdam);  
T. Yanagida, KEK Report No. 79-18, 1979; 
R. Mohapatra and G. Senjanovic,  \prl{44}{1980}{912}.
%
\bibitem{lepto_review}
M. Fukugita and T. Yanagida, \plb{174}{1986}{45};
M. Luty, \prd{45}{1992}{455};
W. Buchm\"uller and M. Plumacher, \plb{389}{1996}{73} and \ijmpa{15}{2000}{5047};
R. Barbieri, P. Creminelli, A. Strumia and N. Tetradis, \npb{575}{2000}{61};
G. F. Giudice, A. Notari, M. Raidal, A. Riotto and A. Strumia, 
\npb{685}{2004}{89};
for a recent review, see W. Buchm\"uller, R. Peccei and T. Yanagida, 
\arnps{55}{2005}{311}.
%
\bibitem{krs} V. Kuzmin, V. Rubakov and M. Shaposhnikov, \plb{155}{1985}{36};
F. Klinkhammer and N. Manton, \prd{30}{1984}{2212}.
%
\bibitem{T_R} W. Buchm\"uller, P. Di Bari and M. Plumacher, \npb{643}{2002}{367}
and Erratum-ibid,{\bf B793} (2008) 362; \annp{315}{2005}{305} and \njp{6}{2004}{105}.
%
\bibitem{wss} H.~Baer and X.~Tata, {\it Weak Scale Supersymmetry: From 
Superfields to Scattering Events}, 
(Cambridge University Press, 2006).
%
\bibitem{gravprob} S. Weinberg, \prl{48}{1982}{1303}.
%
\bibitem{kl} M. Khlopov and A. Linde, \plb{138}{1984}{265}.  
%
\bibitem{ntlepto} G. Lazarides and Q. Shafi, \plb{258}{1991}{305};
K. Kumekawa, T. Moroi and T. Yanagida, \ptp{92}{1994}{437};
T. Asaka, K. Hamaguchi, M. Kawasaki and T. Yanagida, \plb{464}{1999}{12}.
%
\bibitem{covi} W. Buchmuller, L. Covi, K. Hamaguchi, A. Ibarra and T. Yanagida,
\jhep{0703}{2007}{037}
%
\bibitem{covi2}
  L.~Covi, J.~Hasenkamp, S.~Pokorski and J.~Roberts,
\jhep{0911}{2009}{003}.
%
\bibitem{Baer:2010kw}
  H.~Baer, S.~Kraml, A.~Lessa and S.~Sekmen,
  %``Reconciling thermal leptogenesis with the gravitino problem in SUSY models
  %with mixed axion/axino dark matter,''
  JCAP {\bf 1011} (2010) 040.
%
\bibitem{pq} R. Peccei and H. Quinn, \prl{38}{1977}{1440} and \prd{16}{1977}{1791}.
%
\bibitem{ww} S. Weinberg, \prl{40}{1978}{223};
F. Wilczek, \prl{40}{1978}{279}.
%
\bibitem{ksvz} J. E. Kim, \prl{43}{1979}{103};
M. A. Shifman, A. Vainstein and V. I. Zakharov, \npb{166}{1980}{493}.
%
\bibitem{dfsz} M. Dine, W. Fischler and M. Srednicki, \plb{104}{1981}{199};
A. P. Zhitnitskii, \sjp{31}{1980}{260}.
%
\bibitem{pqmssm} 
H. P. Nilles and S. Raby, \npb{198}{1982}{102};
J. E. Kim, \plb{136}{1984}{378};
J. E. Kim and H. P. Nilles, \plb{138}{1984}{150}.
%
\bibitem{axino} For recent reviews of axino dark matter, see
F. Steffen, \epjc{59}{2009}{557}; L. Covi and J. E. Kim, 
\njp{11}{2009}{105003}.
%
%\bibitem{axion} For recent reviews on axion physics, see
%M. Turner, \prep{197}{1990}{67};
%P. Sikivie, \hepph{0509198};
%J. E. Kim and G. Carosi, arXiv:0807.3125.
%
\bibitem{ckn} A. Cohen, D. Kaplan and A. Nelson, 
\plb{388}{1996}{588}.
%
\bibitem{esusy} H. Baer, S. Kraml, A. Lessa, S. Sekmen and X. Tata,
\jhep{1010}{2010}{018}.
%
\bibitem{MU} K. Choi, A. Falkowski, H. P. Nilles, M. Olechowski and
S. Pokorski, \jhep{0411}{2004}{076}; K. Choi, A. Falkowski, H. P. Nilles
and M. Olechowski, \npb{718}{2005}{113}; K. Choi, K-S. Jeong,
  \jhep{0701}{2007}{103}; A. Falkowski, O. Lebedev and Y. Mambrini,
  \jhep{0511}{2005}{034}; H. Baer, E. Park, X. Tata and T. T. Wang, 
\jhep{0608}{2006}{041}; \plb{641}{2006}{447}; \jhep{0706}{2007}{033}.
%
\bibitem{ay} T. Asaka and T. Yanagida, \plb{494}{2000}{297}.
%
\bibitem{wmap7} E. Komatsu {\it et al.} (WMAP collaboration), 
arXiv:1001.4538 (2010).
%
\bibitem{msugra} For a recent review, see
R. Arnowitt and P. Nath, arXiv:0912.2273 (2009). 
%
\bibitem{rtw} K. Rajagopal, M. Turner and F. Wilczek, 
\npb{358}{1991}{447}.
%
\bibitem{ckr} J. E. Kim, A. Masiero and D. V. Nanopoulos, \plb{139}{1984}{346}; 
L. Covi, J. E. Kim and L. Roszkowski, \prl{82}{1999}{4180}.
%
\bibitem{ckkr} L. Covi, H. B. Kim, J. E. Kim and L. Roszkowski, \jhep{0105}{2001}{033}.
%
\bibitem{steffen} A. Brandenburg and F. Steffen, JCAP{\bf 0408} (2004) 008.
%
\bibitem{strumia} A. Strumia, \jhep{1006}{2010}{036}.
%
\bibitem{isared} H. Baer, C. Balazs and A.Belyaev, \jhep{0203}{2002}{042}.
%
\bibitem{isatools} H. Baer, C. Balazs, A.Belyaev, J. K. Mizukoshi, 
X. Tata and Y. Wang, \jhep{0207}{2002}{050}.
%
\bibitem{isajet} F. Paige, S. Protopopescu, H. Baer and X. Tata, \hepph{0312045}; 
http://www.nhn.ou.edu/$\sim$isajet/
%
\bibitem{relic_G} M. Bolz, A. Brandenburg and W. Buchmuller, \npb{606}{2001}{518};
J. Pradler and F. Steffen, \prd{75}{2007}{023509};
V. S. Rychkov and A. Strumia, \prd{75}{2007}{075011}.
%
\bibitem{vacmis} L. F. Abbott and P. Sikivie, \plb{120}{1983}{133};
J. Preskill, M. Wise and F. Wilczek, \plb{120}{1983}{127};
M. Dine and W. Fischler, \plb{120}{1983}{137};
M. Turner, \prd{33}{1986}{889}; L. Visinelli and P. Gondolo, 
\prd{80}{2009}{035024}.
%
\bibitem{axdm} H. Baer, A. Box and H. Summy, 
\jhep{0908}{2009}{080}.
%
\bibitem{jlm} K. Jedamzik, M. LeMoine and G. Moultaka,
JCAP{\bf 0607} (2006) 010.
%
\bibitem{warm}
  M.~Viel, G.~D.~Becker, J.~S.~Bolton, M.~G.~Haehnelt, M.~Rauch and W.~L.~W.~Sargent,
  Phys.\ Rev.\ Lett.\  {\bf 100} (2008) 041304;
  A.~V.~Maccio' and F.~Fontanot,  arXiv:0910.2460 [astro-ph.CO]; 
  A.~Boyarsky, J.~Lesgourgues, O.~Ruchayskiy and M.~Viel, JCAP {\bf 0905} (2009) 012; 
  D.~Boyanovsky and J.~Wu, arXiv:1008.0992 [astro-ph.CO].
%
\bibitem{hot}
  S.~Hannestad, A.~Mirizzi, G.~G.~Raffelt and Y.~Y.~Y.~Wong,
  JCAP {\bf 1008} (2010) 001.
%
\bibitem{ellis} 
R. H. Cyburt, J. Ellis, B. D. Fields and K. A. Olive, \prd{67}{2003}{103521};
R. H. Cyburt, J. Ellis, B. D. Fields, F. Luo, K. Olive and V. Spanos, 
JCAP{\bf 0910} (2009) 021.
%
\bibitem{kohri} M. Kawasaki, K. Kohri and T. Moroi, 
\plb{625}{2005}{7} and \prd{71}{2005}{083502};
K. Kohri, T. Moroi and A. Yotsuyanagi, \prd{73}{2006}{123511};
for an update, see M. Kawasaki, K. Kohri, T. Moroi and A. Yotsuyanagi, 
\prd{78}{2008}{065011}.
%
\bibitem{jedamzik} K. Jedamzik, \prd{70}{2004}{063524}
and \prd{74}{2006}{103509}.
%
\bibitem{stau} J.~Ellis, T.~Falk and K.~Olive, \plb{444}{1998}{367}; 
J.~Ellis, T.~Falk, K.~Olive and M.~Srednicki, \app{13}{2000}{181};
M.E.~G\'{o}mez, G.~Lazarides and C.~Pallis, \prd{61}{2000}{123512}
and \plb{487}{2000}{313};
A.~Lahanas, D.~V.~Nanopoulos and V.~Spanos, \prd{62}{2000}{023515};
R.~Arnowitt, B.~Dutta and Y.~Santoso, \npb{606}{2001}{59}; 
see also Ref.~\cite{isared}.
%
\bibitem{Afunnel} M.~Drees and M.~Nojiri, \prd{47}{1993}{376}; 
H.~Baer and M.~Brhlik, \prd{57}{1998}{567};
H.~Baer, M.~Brhlik, M.~Diaz, J.~Ferrandis, P.~Mercadante,
P.~Quintana and X.~Tata, \prd{63}{2001}{015007};
J.~Ellis, T.~Falk, G.~Ganis, K.~Olive and M.~Srednicki, \plb{510}{2001}{236}; 
L.~Roszkowski, R.~Ruiz de Austri and T.~Nihei, \jhep{0108}{2001}{024}; 
A.~Djouadi, M.~Drees and J.~L.~Kneur, \jhep{0108}{2001}{055}; 
A.~Lahanas and V.~Spanos, \epjc{23}{2002}{185}.
%
\bibitem{hb_fp} K.~L.~Chan, U.~Chattopadhyay and P.~Nath, \prd{58}{1998}{096004};
J.~Feng, K.~Matchev and T.~Moroi, \prl{84}{2000}{2322} and \prd{61}{2000}{075005}; 
see also H.~Baer, C.~H.~Chen, F.~Paige and X.~Tata, \prd{52}{1995}{2746} and 
\prd{53}{1996}{6241}; 
H.~Baer, C.~H.~Chen, M.~Drees, F.~Paige and X.~Tata, \prd{59}{1999}{055014}; 
for a model-independent approach, see
H.~Baer, T.~Krupovnickas, S.~Profumo and P.~Ullio, \jhep{0510}{2005}{020}.
%
\bibitem{kns} M. Kawasaki, K. Nakayama and M. Senami, JCAP{\bf 0803} (2008) 009.
%
\bibitem{kim} J. E. Kim, \prl{67}{1991}{3465}.
%
\bibitem{turner} R. Scherrer and M. S. Turner, \prd{31}{1985}{681};
see also G. Lazarides, R. Schaefer, D. Seckel and Q. Shafi, 
\npb{346}{1990}{193} and J. E. Kim, Ref. \cite{kim};
see also   J.~Hasenkamp and J.~Kersten, arXiv:1008.1740 [hep-ph].
%
\bibitem{epsilon1} W. Buchm\"uller and S. Fredenhagen, \plb{483}{2000}{217}.
%
\bibitem{admx} 
L. Duffy {\it et al.}, \prl{95}{2005}{091304} and \prd{74}{2006}{012006};
for a review, see S. Asztalos, L. Rosenberg, K. van Bibber, P. Sikivie
and K. Zioutas, \arnps{56}{2006}{293}.
%
\bibitem{superiso}
F. Mahmoudi, \cpc{178}{2008}{745}
%
\bibitem{freitas} A. Freitas, F. Steffen, N. Tajuddin and D. Wyler,
\plb{679}{2009}{270} and \plb{682}{2009}{193}. 
%
\bibitem{pdb} K. Nakamura {\it et al.} (Particle Data Group), 
\jpg{37}{2010}{075021}.
%


%%%%%%%%%%%%%%%%%%%%%%%%%%%%%%%%%%%%%

\end{thebibliography}
\end{document}